\documentclass[11pt,a4paper]{article}
\title{Analysis of Forensic DNA Mixtures with Artefacts}
\author{R. G. Cowell\\City University London\and T. Graversen\\University of Oxford\and S. L. Lauritzen\thanks{Department of Statistics, University of Oxford, 1 South Parks Road, Oxford OX1 3TG, United Kingdom.}\\University of Oxford\and
J. Mortera\\Universit{\`a} Roma Tre}
\date{}
\usepackage{natbib,graphicx,amsmath}
\usepackage[margin=12pt, font = small, labelfont =bf]{caption}
\usepackage{xspace,enumerate}

\usepackage{tikz, ifthen}
\usetikzlibrary {shapes, arrows}
\usetikzlibrary {positioning}
\usetikzlibrary{fit, backgrounds, calc, matrix}

\newcommand{\secref}[1]{\mbox{Section~\ref{sec:#1}}}
\newcommand{\figref}[1]{\mbox{Fig.~\ref{fig:#1}}}
\newcommand{\tabref}[1]{\mbox{Table~\ref{tab:#1}}}
\renewcommand{\eqref}[1]{\mbox{(\ref{eq:#1})}}
\newcommand{\Secref}[1]{\mbox{Section~\ref{sec:#1}}}

\newcommand{\Tabref}[1]{\mbox{Table~\ref{tab:#1}}}

\newcommand{\ie}{\emph{i.e.}\xspace}
\newcommand{\cd}{\,|\,}
\newcommand{\B}{{\cal B}}
\newcommand{\E}{{\mbox{E}}}
\newcommand{\hyp}{\mathcal{H}}
\newcommand{\amp}{\mathbin{{\scriptstyle\&}}}

\begin{document}

\maketitle
\begin{abstract}
DNA is now routinely used in criminal investigations and court cases, although DNA samples taken at crime scenes are of varying quality and therefore present challenging problems for their interpretation.
  We present a statistical model for the quantitative peak information
  obtained from an electropherogram (EPG) of a forensic DNA sample and illustrate its potential use for the analysis of criminal cases. In contrast to most previously used methods, we
 directly model the peak height information and incorporates important artefacts associated with the production of the EPG. Our model has a number of unknown parameters, and we show that these can be estimated by the method of maximum likelihood in the presence of
  multiple unknown contributors,  and their approximate standard
  errors calculated; the computations exploit a Bayesian network
  representation of the model.  A case example from a UK
  trial, as reported in the literature, is used to illustrate the efficacy and use of the model, both in
  finding likelihood ratios to quantify the strength of evidence, and in the
  deconvolution of mixtures for the purpose of finding likely profiles
  of one or more unknown contributors to a DNA sample. 
  Our model is
  readily extended to simultaneous analysis of more than one mixture
  as illustrated in a case example. We show that combination of
  evidence from several samples may give an evidential strength close
  to that of a single source trace and thus modelling of peak height
  information provides for a potentially very efficient mixture analysis.
  \end{abstract}
\noindent \textbf{Key words:} allelic dropout; Bayesian networks; DNA profiles; forensic statistics; gamma distribution;  mixture deconvolution; silent alleles, strength of evidence, stutter.
\section{Introduction}
Since the pioneering work of \cite{jeffreys:85}, genetic fingerprinting or DNA profiling has become an indispensable tool for identification of individuals in the investigative and judicial process associated with criminal cases, in paternity and immigration cases, and other  contexts \citep{aitken:taroni:04,balding:05}. The technology has now reached a stage where  uncertainty associated with determining the identity of an individual based on matching a high quality DNA sample from a crime scene to one taken under laboratory conditions has been virtually eliminated.  Frequently, however, DNA samples found on crime scenes are more complex, either because the amounts of DNA are tiny, even just a few molecules, the samples contain DNA from several individuals, the DNA molecules have degraded, or a combination of these. In such cases there is considerable uncertainty involved in determining whether or not the DNA of a given individual, for example a suspect in a criminal case, is actually present in the given sample. A sample  found on the crime scene may contain material from the victim but also from individuals which might be involved in the crime, and determining their DNA becomes an important issue for the criminal investigation. In many cases such \emph{mixed} samples contain DNA from, say, a victim and a perpetrator, but advanced DNA technology now extracts genetic material from a huge variety of surfaces and objects which may have been handled by several individuals, only some of them being related to the specific crime. DNA mixtures with DNA from many individuals occur frequently in multiple rape cases or in traces left by groups of perpetrators handling the same objects such as, for example, balaclavas and crowbars. The basic representation of the composition of a DNA sample is the \emph{electropherogram} (EPG), as described further in \secref{dnaintro} below and displayed in \figref{epg}. 

The identification of the DNA composition of such mixed samples gives rise to a wide range of challenging statistical questions, some associated with uncertainties and artefacts in the measurement processes, some associated with population genetic variations, and other of a more conceptual nature. This article attempts to address some of these challenges. 

  As detailed in \secref{litreview} below, the statistical issues concerning DNA mixtures have previously been addressed in a number of different ways. 
Some approaches make use only of the presence or absence of peaks in the EPG or use the peak heights in a semi-quantitative fashion; other approaches seek to develop fully quantitative models of observed peak heights. Although several methods have been developed, they each suffer from limitations of various kinds and none of the methods have been universally acknowledged as gold standards. 

This article presents a new statistical model for the peak heights of an EPG. It builds upon earlier models presented in \cite{article:Gammamodel, rgc/sll/jm:fsi,Cowell2011202} but is simpler than these, for example
eliminating the need for previously introduced discretizations of some continuous parameters.  
The simplifications of the model combined with  an efficient Bayesian network representation \citep{graversen:lauritzen:comp:13} enables fast computation  and permits analysis of mixtures in which the presence  of several unknown contributors is posited; in particular, estimation of unknown parameters by maximum likelihood becomes computationally feasible.  

The plan of the paper is as follows. In the next section we present  background information concerning DNA relevant for  mixture analyses carried out by forensic scientists. We summarize the measurement processes carried out for quantifying DNA mixtures, and the artefacts that can arise in these processes and lead to difficulties in their interpretation. A case example  from a trial described in the literature is  presented in \secref{pubcase} and used for illustration in the remainder of the paper.  In \secref{objectives} we describe the basic elements associated with statistical interpretation of DNA evidence. In \secref{gammamodel} we give a detailed description of our basic model for peak heights and its elaboration with inclusion of various artefacts associated with the EPG  and we briefly review related work on DNA mixture modelling  in \secref{litreview}; we note that also \cite{steele:balding:14} gives a review of recent literature and software for DNA mixture analysis. In \secref{case} we apply our model to the case presented in \secref{pubcase}, estimating the unknown parameters in our model by maximum likelihood using an efficient Bayesian network representation. \Secref{artefacts} is devoted to showing how the modularity of the Bayesian network representation may be used to 
elicit further details in the analysis of  a mixture, for example by finding the probability that a particular peak is a stutter peak; it is also shown how the networks may be extended for other purposes. In \Secref{single} we show how the model can also be useful for the analysis of DNA from a single individual.
Finally 
we discuss suggestions  for further work in \secref{discuss}.
\subsection{DNA mixtures}
\label{sec:dnaintro}

In this section we describe the nature of the information that  we  analyse with our DNA mixture model and give a brief description of
the DNA amplification process and associated measurements as carried out in forensic  laboratories. We also
introduce some of the complications
that can occur. A more comprehensive description may be found in  \cite{book:Butler}.

\subsubsection{Short Tandem Repeat (STR) markers}

Forensic scientists encode an individual's genetic profile using the composition of DNA at various positions
 on the  chromosomes.

 A specific position on a chromosome is called a
\textit{locus}, or \textit{marker}. The information at each locus consists of an unordered pair of \textit{alleles}  which forms the \textit{genotype} at that locus; a pair because
 chromosomes come in pairs, one inherited from the father and one from the mother, and unordered because
 it is not recorded from which chromosome of the pair each allele originates. We adopt the standard assumption that the population is in Hardy--Weinberg equilibrium implying that the two alleles of an individual can be assumed to be sampled independently from the same population. Human DNA has twenty three pairs of chromosomes:
twenty two autosomal chromosome pairs and a sex-linked pair.

The loci used for forensic identification have been chosen for various reasons. Among these, we point out the following two. The first reason is that at each locus there is a wide  variability between individuals in
the alleles that may be observed. This variability can therefore
 be exploited to differentiate people.
The second reason is that each  locus is either on a distinct chromosome, or if a pair of loci are on the same chromosome then they are widely
separated. This means that the alleles at the various loci may be treated as mutually independent, thus simplifying the statistical analysis.

The alleles of a marker are sequences of the four amino acid nucleotides \emph{adenine, cytosine, guanine}  and \emph{thymine}, which we  represent by the
  letters A, C, G and T. Each amino acid is also called a \textit{base}, and because the DNA molecule has a double helix structure,
 each amino acid on one strand is linked to a complementary amino acid on the other strand; a complementary
 pair of amino acids is called a \textit{base pair}.

An allele is typically named by its \textit{repeat number}, usually an integer. For example,  consider the allele with
repeat number 5 (commonly also referred to as allele 5 for brevity) of the marker TH01.  This allele includes the sequence of four nucleotides AATG
repeated consecutively five times. It can be designated by the formula $[AATG]_5$. Likewise allele 8 of TH01 has eight consecutive repetitions of
the AATG sequence, which may be denoted by $[AATG]_8$. Repeat numbers are not always integers. For example, allele 8.3 of TH01 has the
chemical sequence $[AATG]_5ATG[AATG]_3$, in which `8' refers to the eight complete  four-word bases $[AATG]$ and the `.3' refers to the three
base-long word sequence  $ATG$ in the middle. Repeat numbers with decimal `.1' and `.2' endings are also possible, indicating the presence of a
word of one or two bases. Note that the integer part of the repeat number counts how many complete words of four bases make up the
allele sequence but the words need not be all identical and may vary even within loci. For example, allele 11 of the marker VWA has the base sequence $TCTA[TCTG]_3[TCTA]_7$. Also, some markers are based on tri- or pentanucleotide motifs rather than tetranucleotides as above. The
base-letter sequences for many alleles may be found in \cite{book:Butler}.

When the repeat numbers of the two alleles of an individual at a marker are the same,
 then the genotype for that marker is said to be \emph{homozygous}; when the repeat numbers differ,
 the genotype for that marker is said to be \emph{heterozygous}.

The repetitive  structure in the alleles gives rise to the term  \emph{short tandem repeat} (STR) marker to describe these loci;
they also go by the name of \textit{microsatellites}.
Note that for other purposes of genetic analysis it is common to use single-nucleotide polymorphisms (SNPs) which are defined as DNA sequence variations that occur when a single nucleotide (A, T, C, or G) in the genome sequence is altered. However, for a number of reasons, the use of STR alleles is dominating in forensic genetics.

Within a population the various alleles of STR markers do not occur equally often, some can be quite common and some quite rare.
When carrying out probability calculations based on DNA, forensic scientists use estimates of probabilities based on allele frequencies in profiles of a sample of
individuals. The sample sizes typically range from a few hundred to thousands of individuals.
For example, \cite{butler:etal:03} presents tables of US-population STR-allele frequencies for Caucasians, African-Americans, and Hispanics based on sample sizes of 302, 258 and 140 individuals. 

\subsubsection{The PCR process}

The DNA collected from a crime scene for forensic analysis consists of a number of human cells from one or more individuals. Note that each cell of an individual will contain two alleles (diploid cells) for each autosomal marker, whereas sperm cells have only one allele (haploid cells).
This means that in a mixture,   a particular individual will contribute the same  number of alleles
for each marker. In order to identify the alleles that are present, a DNA sample is first subjected to  chemical reagents that break down the cell walls
so that the individual  chromosomes are released into a solution. A small amount of this  solution is exploited to quantify the concentration of DNA;  the typical unit of measurement is  picograms per microlitre, the DNA in a single human cell having a mass of around  six picograms. Having determined the density of DNA in the sample, a volume is extracted that is estimated to contain a certain quantity of DNA, typically around 0.5 nanograms, equivalent to around 100  human cells.
The DNA in the extract is then amplified using the \emph{polymerase chain reaction} (PCR) process. This involves  adding \textit{primers} and other biochemicals to the extract, and then subjecting it to a number of rapid heating and cooling cycles. Heating the extract has the effect of splitting apart
the two complementary strands  of DNA, the cooling phase then  allows free floating amino acids to bind with these individual strands in such a way that the DNA is copied. By the action of repeated heating and cooling cycles, typically around 28 altogether, an initially small amount
of DNA is amplified to an amount large enough for quantification. Mathematically, the amplification process may be modelled as a branching process \citep{article:SunPCR,article:StolovitzkyCecchi}. The amplification process is not 100\% efficient, that is, not every allele gets copied in each cycle. This means that if two distinct alleles in a marker are present in the extract in the same amount prior to amplification, they will occur in different amounts at the end of the PCR process.
In our model we capture this variability using  gamma distributions.

Note that after breaking down the cell walls to release the chromosomes there could be  sufficient DNA in the sample to carry out PCR amplifications with several extracts. When this is done it is called a \textit{replicate run}.

To understand the quantification stage of the post PCR amplified DNA, it is important to know that the amplification process does not copy only the
repeated DNA word segment of a marker, it also copies DNA at either end. These are called \textit{flanking regions}, and their presence is important
in performing the PCR process. Thus an amplified allele will consist of the allele word sequence itself and two flanking regions, and will have a length associated with it which is measured in the total number of base pairs included in the word sequence and the flanking regions. For each marker, the DNA sequence (hence the size) forming each of the two flanking regions is constant, but different across markers. Thus quantifying a certain allele is equivalent to measuring how much
DNA is present of a certain size. This is carried out by the process of \textit{electrophoresis}, as follows.

\begin{figure}[htb]
  \begin{center}
     \includegraphics[width=0.95\textwidth]{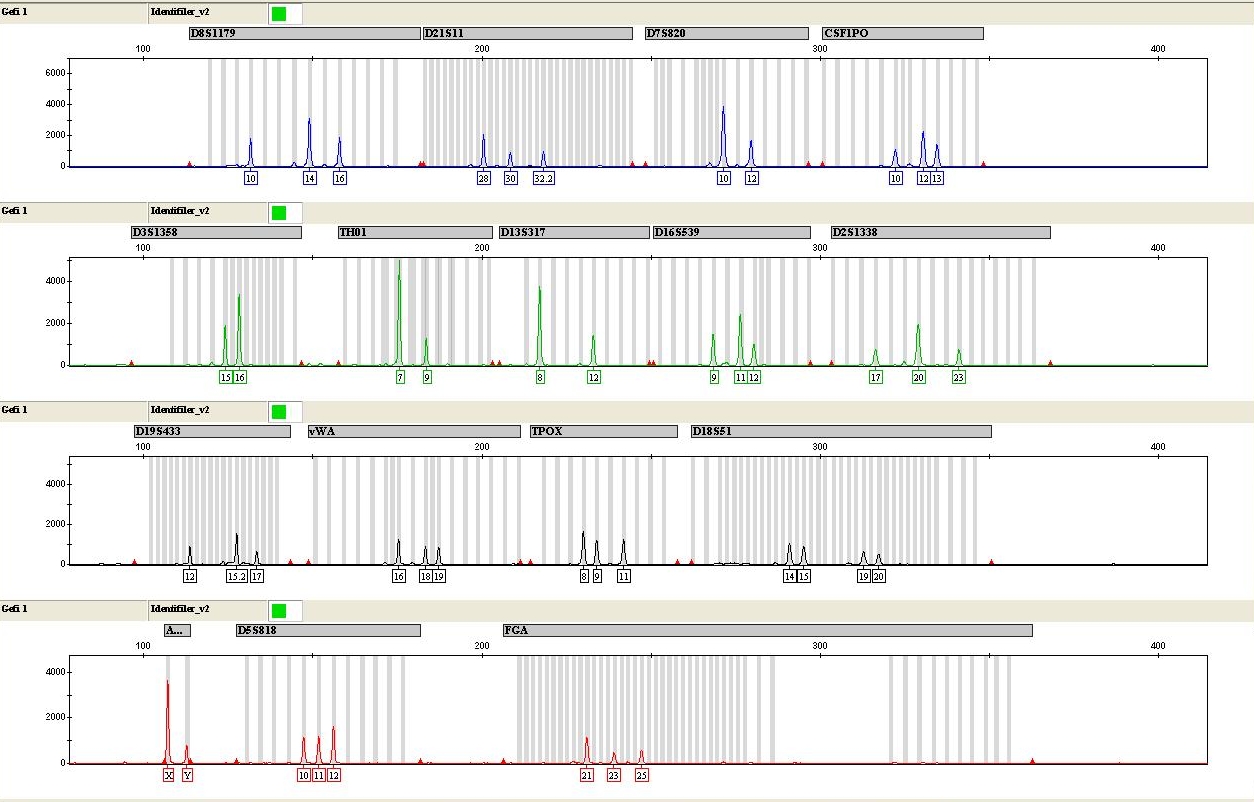}
  \caption{An electropherogram (EPG) using the Identifiler\textsuperscript{TM} STR kit. Each of the four panels represents a dye.
Marker names are in grey boxes above panels and repeat numbers of detected alleles are given below the corresponding peaks. The vertical grey bars are used to identify alleles by their length in base pairs. On the left-hand side of each panel the scale of RFU is indicated.}
  \label{fig:epg}
  \end{center}
\end{figure}

The flanking regions have attached to them a fluorescent dye. Several colours of fluorescent dyes are used to distinguish 
similarly sized alleles from different markers.
The amplified DNA is drawn up electrostatically through a fine capillary to pass through a
light detector, which illuminates the DNA with a laser and measures the amount of fluorescence generated. The latter is then an indication of the number of alleles tagged with the fluorescent marker. The  longer  alleles are drawn up more slowly than the shorter alleles, however alleles of the same length are drawn up together. This means that the intensity of the detected fluorescence will sharply peak as a group of alleles of the same length passes the light detector, and the value of the intensity will be a measure of the number of alleles that pass. The detecting apparatus thus measures a time series of fluorescent intensity, but it  converts the  time variable into an equivalent
base pair length variable. The data may be presented to a forensic scientist as  an \emph{electropherogram} (EPG) as shown in  \figref{epg} with each  panel in the EPG corresponding to a different dye. The  horizontal axes indicate the base pair length, and the vertical axis the intensity.

In the absence of artefacts,  a peak in the EPG indicates  presence of an allele in the sample before amplification. The peak height is
a measure of the amount of the allele in the amplified sample expressed in \textit{relative fluorescence units}
(RFU). The area of the peak is another measure of the amount, but
this is highly correlated with the height \citep{Tvedebrink2010}. Both
peak height and peak area are determined by software in the detecting
apparatus. 

 We shall call the peak size information
extracted from the EPG the \textit{profile of the DNA sample}, or more briefly,
the {DNA profile}. Commonly, DNA profile also refers to the combined genotype of a person across all markers.

In measuring  the peak heights, low level noise give rise to small spurious peaks. A \textit{peak amplitude threshold} may be
set by the forensic analyst whereby peaks below the threshold level are ignored.  Thus an allele present in the DNA sample will not be
recorded as observed if the peak it generates is below the threshold; when this happens a \textit{dropout} of the allele is said to have
occurred. A dropout can also occur simply by sporadic failure of the apparatus.
Dropout is an artefact that can make the analysis of DNA samples difficult. Another common artefact is
\textit{stutter}, where\-by an allele that is present in the sample is mis-copied at some stage in the PCR amplification process, and a four base pair
word segment is omitted; more rarely other stutter patterns occur. This damaged copy  itself takes part in the amplification process, and so yields a peak
four base pairs below the allele from which it arose.

Another artefact is known as \textit{dropin}, referring to the occurrence of  small unexpected peaks in the EPG. This can for example be due to sporadic contamination of a sample either at source or in the forensic laboratory.

Current technology allows for the amplification of very little DNA material, even as little as contained within one cell \citep{Findlay1997}. Amplifications of low
amounts of DNA are termed low template DNA (LTDNA).  LTDNA analyses  are particularly prone to artefacts such as dropin and dropout.

Finally, a mutation in the flanking region  can result in the allele not being picked up at all by the PCR process, in which case we say that the allele is \emph{silent}. An allele can also be undetectable and thus \emph{de facto} silent because its length is off-scale and the peak does therefore not appear in the EPG.

\subsection{Motivating example}
\label{sec:pubcase}
As a motivating example we shall consider a case from a UK trial as reported in  \cite{Gill200891}, also analysed in \cite{Cowell2011202}:
\begin{quote}``An incident had occurred in a public house where the
deceased had spent the evening with some friends. There was an
altercation in the car park between the deceased ($K_1$) and
several others resulting in the death of the victim. The alleged
offenders then left the scene and went to another public house
where they were seen to go into the lavatory to clean
themselves." \end{quote}

Two blood stains, MC18 and MC15, were found at the public house lavatory. Both indicated that they were DNA mixtures of at least three
individuals. The genotypes of the \emph{defendant, $K_3$}, and a known individual, $K_2$, alleged to be present at the time of the offence,
were determined together with that of the \emph{victim $K_1$} (all were males).  An excerpt of the observed DNA profiles of the samples and
individual genotypes are displayed   in \tabref{mc1518dat}. The complete data  can be found in \cite{Gill200891} and in the R-package \texttt{DNAmixtures} \citep{graversen:package:13}.
\begin{table}
\caption{\label{tab:mc1518dat}Alleles, peak heights, and genotypes of  individuals for an excerpt of the markers in the pub crime case.}
  \centering\small
\begin{tabular}{lrrrrrr}
 Marker &Alleles & MC18& MC15& $K_1$& $K_2$ & $K_3$\\
 \hline
 D2 & 16 & 189 & 64 &  & & 16   \\
      & 17 & 171 & 96 &  & & 17      \\
      & 22& 55 & 0&  & & \\
      & 23 & 638 & 507 & 23 & & \\
      & 24 & 673& 524 & 24 &24 & \\
     \hline
 D16 & 11 & 534& 256 &   & 11 &  11  \\
      & 12 & 1786 & 1724 & 12 & 12&    \\
      & 13 & 265 & 109 &   & &  13    \\
      \hline
 TH01 & 7 & 670     & 727   & 7 &  &    \\
      & 8 & 636    & 625   & 8 &  &     \\
 & 9 & 99&   0 &   & 9 &     \\
      & 9.3 & 348 &165     &   &  & 9.3\\
      \hline
\end{tabular}

\end{table}

Note that if the trace consists of DNA from exactly $K_1$, $K_2$,  and $K_3$, the peak at allele 22 of marker D2 in MC18 would need to be due to stutter and/or spontaneous dropin, and $K_2$'s allele 9 on marker
TH01 would have dropped out of the  MC15 profile. We emphasize that the entire profile would be consistent with only DNA from $K_1$, $K_2$,  and $K_3$ being present in the traces. 

\subsection{Objectives of analysis}
\label{sec:objectives}
The analysis of a  DNA profile can have different objectives  depending on the context. The objective can be a quantification of the strength of
evidence for a given hypothesis over another, or the objective may be a \emph{deconvolution} of the profile, \ie  one wishes to identify likely
genotypes of contributors \citep{article:Perlindata,wang2006least,tvedebrink:deconv:12}. We briefly describe the typical situations below.
 \paragraph{Weight of evidence}
The available \emph{evidence} $E$  consists of the peak heights as observed in the EPG as well as the combined genotypes of the known individuals. It is customary to assume relevant
population allele frequencies to be known. We shall return to this issue in \secref{discuss}.

To quantify the strength of the evidence against the defendant ($K_3$), two competing hypotheses are typically specified. One of these, usually referred to as the \emph{prosecution hypothesis} $\hyp_p$, could in our example be that the profile has exactly three contributors who are
identical to the three individuals $K_1$, $K_2$, and $K_3$. Since their genotypes are known, we refer to these as \emph{known contributors}. Alternatively, $\hyp_p$ could only involve $K_1$ and $K_3$ as known contributors in addition to 
one or more unknown
individuals who we shall name $U_1, U_2, U_3,\ldots$ and refer to as \emph{unknown contributors}. The alleles of the unknown contributors are
assumed to  be chosen randomly and independently from a reference population with known allele frequencies. To limit the scope of our analysis we shall here only consider scenarios which include $K_1$ and $K_2$ as contributors to the traces.

The prosecution hypothesis is then compared to what is referred to as the \emph{defence hypothesis} $\hyp_d$, which typically would replace $K_3$ in $\hyp_p$ with an unknown contributor, claiming that the apparent similarity between the trace and the DNA profile of $K_3$ is due to chance. The defence hypothesis is not actually advocated by the defence, but is formulated strictly for comparative purposes. 
We emphasize that this implies that the genotypes of this unknown contributor in principle could be identical to that of $K_3$ although this would be extremely improbable, ignoring cases involving identical twins. 

 The strength of the evidence \citep{good1950,lindley1977,balding:05}  is normally 
represented by the \emph{likelihood ratio}:
$$ LR=\frac{L(\hyp_p)}{L(\hyp_d)}=\frac{\Pr( E \cd \hyp_p)}{\Pr(E \cd \hyp_d)}.$$
We shall follow \cite{balding:13} and report the \emph{weight of evidence} as $\mbox{WoE} =\log_{10}LR$ in  the unit \emph{ban} introduced by Alan Turing \citep{good:79} so that one ban represents quality,
a factor 10 on the likelihood ratio. For a high quality single source DNA profile (\textit{i.e.} from a single individual)  in the SGM Plus system, values for WoE vary around 14 bans \citep{Balding:99}, which in most cases is sufficient for establishing the identity of the perpetrator beyond reasonable doubt. We emphasize that this unit of WoE  is not absolute, but should be always be interpreted relative to a specific defence hypothesis; it could for example change --- in principle in any direction --- if also $K_2$ were replaced by an unknown individual.

The calculation of the WoE can be important both when presenting a case to the court and in the investigative phase of a trial, to decide whether it is worthwhile to search for additional independent evidence.

The numerator and denominator in the likelihood ratio are calculated based on models which shall be detailed further in \secref{gammamodel} below; generally models have the form
$$\Pr( E \cd \hyp)= \sum_\mathbf{g}\Pr( E \cd \mathbf{g}) \Pr( \mathbf{g} \cd \hyp)$$
so that the model for the conditional distribution $\Pr( E \cd \mathbf{g})$ of the evidence given the genotypes $\mathbf{g}$ of all contributors  is the same for both hypotheses, whereas the hypotheses differ concerning the distribution $\Pr( \mathbf{g} \cd \hyp)$ of genotypes of the contributors, as described above. 
\paragraph{Deconvolution of DNA mixtures}
This calculation attempts to identify  the combined genotypes across all markers of each unknown
contributor to the mixture and give a list of potential genotypes of a perpetrator to use for a database search.
For example, we could wish to calculate
$$\Pr\{U_1,U_2\cd  E, \hyp_{\rm{inv}}\} \mbox{ or } \Pr\{U_1\cd  E, \hyp_{\rm{inv}}\},$$
where $U_1,U_2$ represent genotypes of  unknown contributors to the mixture under an investigative hypothesis $\hyp_{\rm{inv}}$ that  the trace contains DNA from three persons, one of whom is  the victim $K_1$, or, for example, two of whom are $K_1$ and $K_2$. The calculation should be done for a range of
probable combinations of genotypes of the unknown contributors.

\paragraph{Evidential efficiency}
For a single source high quality trace the WoE specializes to $-\log_{10}$  of the \emph{match probability}   i.e.\ the probability  that a random member of the population has the  specific DNA profile of $K_s$ \citep{balding:05}.
We  point out that the WoE against a suspect $K_s$ based on our model can never be stronger than the WoE obtained by a single source DNA profile. 
To see this, let $\pi_s= \Pr(U=K_s)$ denote the match probability. Further, let $\hyp_p$ be the prosecution hypothesis involving $K_s$ as a contributor and $\hyp_d$ the defence hypothesis, replacing the genotype $K_s$ by that of a random individual $U$.
We then have
\begin{eqnarray}LR=\frac{\Pr(E\cd \hyp_p) }{\Pr(E\cd \hyp_d)}&= &
\frac{\Pr(E\cd \hyp_p) }{\Pr(E\cd \hyp_d, U=K_s)\pi_s +(1-\pi_s)\Pr(E\cd \hyp_d, U\neq K_s)} \nonumber \\&=&
\frac{\Pr(E\cd \hyp_p) }{\Pr(E\cd \hyp_p)\pi_s +(1-\pi_s)\Pr(E\cd \hyp_d, U\neq K_s)}\nonumber  \\&\leq &
\frac{\Pr(E\cd \hyp_p) }{\Pr(E\cd \hyp_p)\pi_s}=\frac{1}{\pi_s}. \label{eq: inequality}
\end{eqnarray}
Thus we have $\mbox{WoE}=\log_{10}LR\leq -\log_{10} \pi_s$ implying that a mixed trace can never give stronger evidence than a high-quality trace from a single source.  
We define the \emph{loss of evidential efficiency} $\mbox{WL}(E\cd K_s)$ against $K_s$ in the evidence $E$  as
\[\mbox{WL}(E\cd K_s)= \mbox{WoE}(K_s)_{\max}-\mbox{WoE}(K_s)=
- \log_{10}\pi_s - \log_{10}\frac{\Pr(E\cd \hyp_p)}{\Pr(E\cd \hyp_d)}.
\]

This quantity is non-negative, and indicates how many bans of WoE are lost due to the evidence being based on a mixture rather than a single source trace. 
Indeed, the loss of efficiency for a mixed trace can be compared to the loss induced by failure of identifying the genotype for some markers, as this increases the match probability in a similar way. For the case example in \secref{pubcase}, using the US Caucasian allele frequencies of  \cite{butler:etal:03}, the negative logarithm of the  match probability for $K_3$  is $-\log_{10}\pi_{K_3}=14.5$, whereas if we ignored, say, markers D18 and D19, the weight-of-evidence would be $\mbox{WoE}=-\log_{10}\tilde\pi_{K_3}=12.0$, losing about 2.5 bans.

In the case of deconvolution without a
specified suspect, we could consider the evidence against the posterior most likely unknown person $K^*$ under the defence hypothesis. If  $\hyp_p^*$ denotes a prosecution
hypothesis obtained by replacing $U$ with $K^*$ and 
 $\pi^* = \Pr(U= K^*) = \Pr(U= K^*\cd \hyp_d)$
 denotes the prior probability of a random individual having genotype $K^*$, we  get
\begin{eqnarray}\mbox{WL}(E\cd K^*)&=& -\log_{10}{\pi^*} - \log_{10}\frac{\Pr(E\cd \hyp_p^*)}{\Pr(E\cd \hyp_d)} \nonumber =
 -\log_{10}\frac{\Pr(E\cd \hyp_p^*)\pi^*}{\Pr(E\cd \hyp_d)} \nonumber
\\& = &-\log_{10}\frac{\Pr(E\cd  U= K^*, \hyp_d)\Pr(U= K^*\cd \hyp_d)}{{\Pr(E\cd \hyp_d)}} \nonumber \\&=&-\log_{10}\Pr(U= K^*\cd \hyp_d, E). \label{eq:genericeff}\end{eqnarray}
Thus, this \emph{generic loss of evidential efficiency} is equal to $-\log_{10}$ of the maximum posterior probability obtained in the deconvolution. For a single source trace that uniquely identifies the contributor, the generic loss of evidential efficiency is 0.

\section{A gamma model with artefacts}
\label{sec:gammamodel} In this section we present an overview of the basic model for peak heights, which is based on the gamma model described in
\cite{article:Gammamodel} and used in \cite{Cowell2011202} with extensions describing artefacts. It should be emphasized that the model used here
is different from the latter in two main respects: firstly it uses absolute peak heights instead of relative peak heights which enables direct
treatment of dropout by thresholding; secondly, it  has a  simpler description of stutter. As a consequence, computations are
greatly simplified so analysis is possible for a high number of unknown potential contributors.

\subsection{Basic model}
We consider  $I$ potential
contributors to a DNA mixture. Let there be $M$ markers used in
the analysis of the mixture with marker $m$ having $A_m$ allelic
types, $m=1,\ldots,M$. Let $\phi_i$ denote the \emph{fraction} of DNA
from individual $i$ prior to PCR amplification, with
$\phi=(\phi_1,\phi_2,\ldots,\phi_I)$ denoting the 
fractions from all contributors. Thus $\phi_i\geq 0$ and $\sum_{i=1}^I \phi_i = 1$. For identifiability of the parameters we further assume that the fractions $\phi_i$ for unknown contributors are listed in non-increasing order, \ie $\phi_a\geq \phi_b$ if $a <b$ are both unknown individuals. It
is assumed that these pre-amplification fractions of DNA are constant
across markers.

For a specific marker $m$ and allele $a$,  the model is describing the total \textit{peak
  height} $H_{a}$.  Ignoring artefacts for the moment,  the model makes
the following further assumptions:
Each contribution
$H_{ia}$ from an individual $i$ to the peak height at allele $a$ has a
gamma distribution, $H_{ia}\sim\Gamma(\rho\phi_in_{ia},\eta)$,
where $\Gamma(\alpha,\beta)$ denotes the distribution with density
$$f(h\cd \alpha,\beta)=\frac{ h^{\alpha-1}}{\Gamma(\alpha)\beta^\alpha}e^{-h/\beta} \mbox{   for $h>0$}.$$
For $\alpha=0$, $\Gamma(0,\beta)$ is the distribution degenerate at $0$. Here $\rho$ is proportional to the total amount of DNA in the mixture prior to
amplification;
the number of alleles of type $a$
  carried by individual $i$ is denoted by $n_{ia}$; and the parameter
$\eta$ determines the scale. We note that the individual contributions $H_{ia}$ are unobservable. Indeed, when artefacts are added to the model, the peak heights $H_{a}=\sum_i H_{ia}$ themselves are unobserved; they will be modified by stutter and dropout as described below in \secref{artef}.

For notational simplicity we have suppressed the dependence of the heights $H^m_{ia}$  and other quantities on markers $m$. It is important for the model that $\phi_i$
are the same for all markers, whereas one would typically expect other parameters to be marker
dependent and possibly dependent on fragment length in case of degraded DNA \citep{Tvedebrink201297}.
 We note that the assumption of a common scale parameter $\eta^m$ for the individual contributions $H^m_{ia}$ for different alleles $a$ enables a simple interpretation of the model in terms of adding independent contributions to peaks both from proper alleles and from stutter, see below.

It becomes practical to introduce the \emph{effective number $B_a(\phi,\mathbf{n})$ of alleles of type $a$}  where
$B_a(\phi,\mathbf{n}) = \sum_i \phi_in_{ia},$ and $\mathbf{n}=(n_{ia}, i=1,\ldots  I; a =1,\ldots, A)$;
we note that as $\sum_i\phi_i=1$, we have that $\sum_a B_a(\phi, \mathbf{n})$ is constant over markers and equal to two in the diploid case.  The term indicates that the peak at allele $a$ behaves as it would for a single contributor with $B_a(\phi,\mathbf{n})$ alleles of type $a$. If the contributor fraction $\phi$ is interpreted as a probability distribution, it is the expected number of alleles of type $a$.
 
 As a sum of independent gamma distributed contributions, the peak height $H_{a}$ at allele  $a$ is gamma distributed as
\begin{equation}\label{eq:gammamodel}H_{ a} \sim \Gamma \{\rho B_a(\phi,\mathbf{n}), \eta\}.\end{equation}
The distribution of $H_{a}$ has expectation $\rho\eta B_a(\phi,\mathbf{n})$ and variance $\rho\eta^2 B_a(\phi,\mathbf{n})$. Thus  the mean of $H_{ a}$  is approximately proportional to the amount of DNA of
  type $a$ and the variance is proportional to the mean. We let  $\mu=\rho\eta$ and $\sigma= 1/\sqrt{\rho}$. Then --- in a trace with only one heterozygous diploid contributor and no artefacts  ---  $\mu$ is the \emph{mean peak height} and $\sigma$ the \emph{coefficient of variation} for peak heights; hence $\sigma$ becomes a measure of generic peak imbalance. In the presentation of results we shall often  use $(\sigma,\mu)$ instead of $(\rho,\eta)$ because of their more direct interpretability.

\subsection{Incorporating artefacts}
\label{sec:artef}
The model described above does not incorporate a number of important artefacts as described in \secref{dnaintro}. This section is devoted to the
necessary modifications needed to take these into account. 

\paragraph{Stutter}
We represent stutter by decomposing  the  individual contributions $H_{ia}$ to peak heights in the EPG into  $H^s_{ia}$ and $H^0_{ia}$ so that $$H_{ia}= H^s_{ia}+H^0_{ia}$$ 
where $H_{ia}^s$ represents   the DNA originating from individual $i$ of allelic type $a$
 that stutters to $a-1$ by losing a repeat number, and $H^0_{ia}$ represents the remainder that goes through the PCR process
undamaged.   We let the components be independent and gamma distributed as
$$H^s_{ia}\sim \Gamma\{ \rho\xi\phi_in_{ia},\eta\}, \quad H^0_{ia}\sim \Gamma\{ \rho(1-\xi)\phi_in_{ia}, \eta\},
$$  where  $\xi$ is the \emph{mean stutter proportion}. The total peak height observed at allele $a$ is then
$$H_{ a}=\sum_i H^0_{ia}  +  \sum_i H^s_{i, a+1} =H^0_a + H^s_{a+1}.$$  Note that now $H_a$  has a slightly different meaning than in  \eqref{gammamodel} above as it has potentially received stutter from the allele above and lost stutter to the allele below. We ignore the possibility that $H_{a}$ might have tiny stutter contributions $H^s_{a+2},H^s_{a+3},\cdots$ from peaks  at $a+2, a+3,\ldots$. In principle, including this is straightforward, but it has strong effects on the complexity of computations \citep{graversen:lauritzen:comp:13}. The new peak height $H_{a}$ is also gamma distributed as
$$H_{a} \sim \Gamma \{\rho D_a(\phi,\xi,\mathbf{n}),\eta\}$$ where now
\begin{equation}\label{eq:Ds}D_a(\phi,\xi, \mathbf{n})=
(1-\xi)B_a(\phi,\mathbf{n})+ \xi B_{a+1}(\phi,\mathbf{n})
\end{equation}
 are the \emph{effective allele counts after stutter.} 
The relative contribution lost to stutter
$$X_a= \frac{H^s_{a}} {H^s_{a}+H^0_{a}}$$ thus follows a beta distribution
$\B\{\xi \rho B_a(\phi,\mathbf{n}), (1-\xi)\rho B_a(\phi,\mathbf{n})\}$ with mean $\E(X_a) = \xi$.  

Notice that our use of  stutter proportion $X_a$ differs from standard practice which uses the ratio 
between the stutter peak and the parent peak \citep{book:Butler}. A peak in stutter position may itself include contributions from proper
alleles, and the parent peak may itself include stutter contributions from an allele with higher repeat number.
Even in the simple event that a peak is entirely due to stutter from a parent peak, which has not itself been inflated by stutter, we have $H_{a}= H^0_{a}$ and $H^s_a = H_{a-1}$ and thus our 
$X_a   =  H_{a-1}/(H_{a-1} + H_{a})$
rather than the conventional ${H_{a-1}} /{H_{a}}$.

\paragraph{Dropout}
The next step in our model is concerned with the fact that in mixtures some alleles are not observed; either because no peak can be identified, or because the peak is below some detection threshold. This phenomenon is prominent for small amounts of DNA.  Here we  take the consequence of the
model and represent dropout by a total peak height  below the chosen  threshold $C$, as described in \secref{dnaintro}. In other
words, we do not observe the height $H_{a}$ as just described, but rather $Z_a$, where
$$Z_a= \left\{ \begin{array}{cr}H_{a} & \mbox{ if $H_{a}\geq C$}\\0&\mbox{otherwise.}\end{array}\right.$$ This implies that the probability that a specific allele is not observed is
\[
P(Z_a=0\cd \mathbf{n})= G\{C;\rho D_a(\phi,\xi,\mathbf{n}),\eta\},\] where
$G$ denotes the cumulative distribution function  of the gamma distribution.

Allellic dropout for single source traces was studied by \cite{Tvedebrink2009,Tvedebrink2012}  who fitted  a logistic regression model to the dropout probability as
\begin{equation}\label{eq:tvededropout}
P(Z_a=0\cd \bar h) = \frac{\alpha \bar h^\beta}{1+\alpha \bar h^\beta}\end{equation}
where $\bar{h}$ is an average of observed peak heights above threshold calculated across all markers. They found that one could use the same value of
$\beta=-4.35$ for all markers, whereas $\alpha$ was marker dependent. The independent variable $\bar h$ was used as a proxy for total amount of DNA.

\begin{figure}[htb]
\begin{center}
\includegraphics{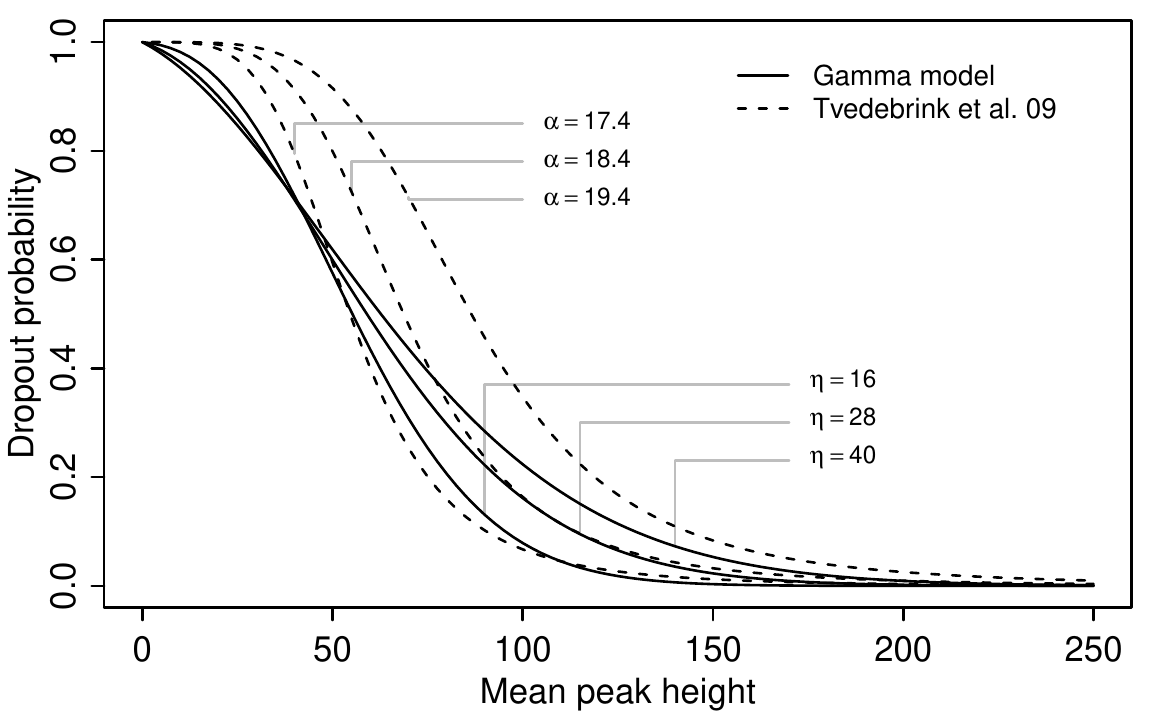}
\end{center}
\caption{\label{fig:dropprob}The probability of dropout of a single allele as a function of mean peak height.  The curves with full lines correspond to our gamma model whereas the dashed curves correspond to the logistic model.}
\end{figure}

In our model, the theoretical mean peak height $\mu=\rho\eta$ would be a similar 
proxy for the total amount of DNA. To compare the two models, we assume that $\bar h $ approximately corresponds to the mean peak height $\mu $.  Thus our model has
\begin{equation}\label{eq:gammacurve}P(Z_a=0\cd \mu)= G\{C;\mu/\eta,\eta\},\end{equation}
which can then be compared with \eqref{tvededropout} where $\bar h$ is replaced by $\mu$.
   A selection of  curves  \eqref{gammacurve} for $C= 50$ and values of $\eta$ corresponding to the maximum likelihood estimate and upper and lower 99\% confidence limits in our case example, see \secref{case}; together with   
curves \eqref{tvededropout}  for $\beta=-4.35$  and representative values of $\alpha$, are superimposed in \figref{dropprob}. We note that the dropout model based on the gamma distribution tends to have lower dropout rates than the logistic model for small amounts of DNA.

\begin{figure}[htb]
\begin{center}
{ \includegraphics{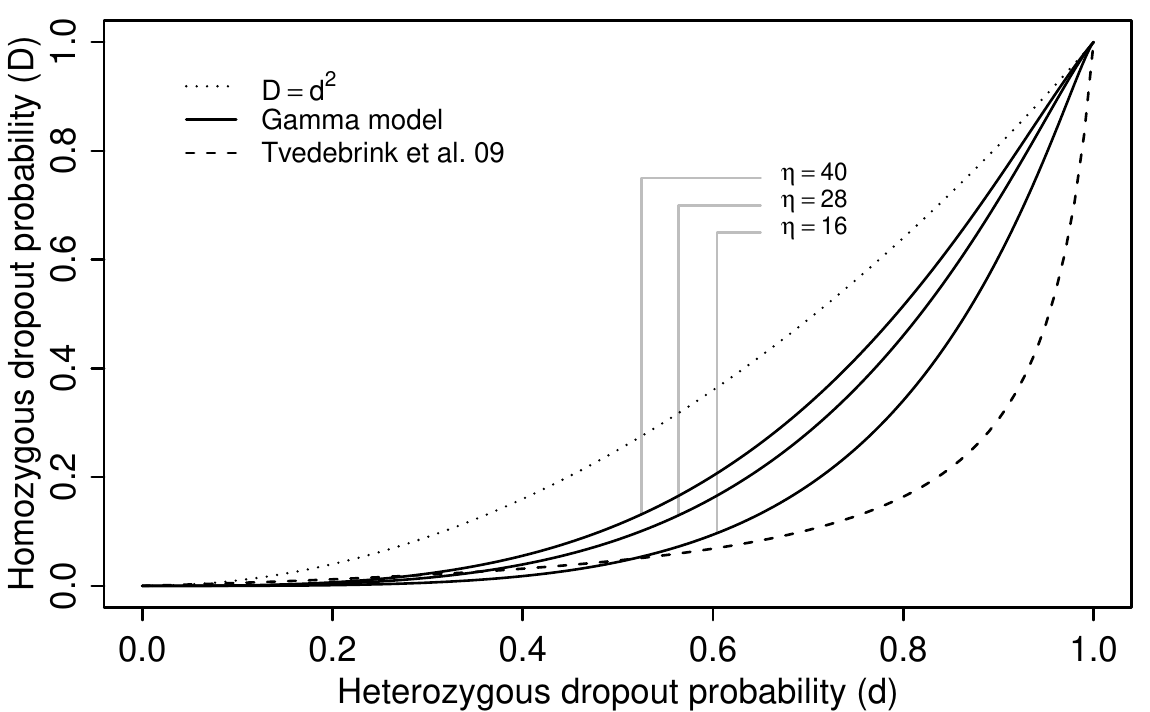}}
\end{center}
\caption{\label{fig:homdrop}The probability of homozygous dropout $D$ as a function of the dropout probability of a single allele for the gamma model and for the logistic model of \cite{Tvedebrink2009}. The solid curves correspond to the gamma model for different values of $\eta$. }
\end{figure}

Both dropout models implicitly define a relation between the dropout probability $d$
for a single allele and the corresponding dropout probability $D$ for a homozygote. \cite{balding:buckleton:09} argue
that these should satisfy $D< d^2$. This relation is satisfied by any threshold model --- hence also for our gamma model --- since if peak height contributions $Y_1$ and $Y_2$ from each allele are independent and identically distributed, we have $$D=P(Y_1+Y_2<C)<P(Y_1<C)^2=d^2.$$ For the model \eqref{tvededropout} 
we have
$$\frac{d}{1-d}=\alpha {\bar h}^\beta;\quad \frac{ D}{1-D} =\alpha (2\bar h)^\beta = 2^\beta \frac{d}{1-d}$$
and thus
\begin{equation}\label{eq:tvedehomdrop}D=\frac{2^\beta d}{1+(2^\beta -1)d}
\end{equation}
independently of the value of $\alpha$.  The curve in \eqref{tvedehomdrop} is displayed in \figref{homdrop} for $\beta=-4.35$ together with the upper bound $D=d^2$ and a selection of curves from the gamma model with $C=50$ and the same values for $\eta$ as in \figref{dropprob}. We note that although \eqref{tvedehomdrop} actually just crosses the curve $D=d^2$ for very small values of $d$, this has hardly any practical significance, as also pointed out in \cite{Tvedebrink2012}. 

\paragraph{Other artefacts}
We have chosen to represent spontaneous dropin of alleles by additional unknown contributors with very small amounts of DNA, implying that most of their alleles will drop out with high probability. With the  computational methods described in  \cite{graversen:lauritzen:comp:13} and  \secref{compute}, this representation is amply feasible. Also, the possibility of a silent allele is easily incorporated into the model, simply by adding an allele that never results in an observed peak, see \secref{artefacts} for details. Note that we have modelled dropout entirely as a threshold phenomenon, ignoring the fact that apparatus failure could be an alternative explanation.

\subsection{Joint likelihood function}
\label{sec:likfun}
For given genotypes of the contributors $\mathbf{n}$, given proportions $\phi$, and given values of the parameters $(\rho,\xi,\eta)$, all observed
peak areas are independent. Thus, the conditional likelihood function based on the observations  $\mathbf{z}=\{z_{ma}, m=1,\ldots, M; a=1,\ldots, A_m\}$ for
all markers $m$ and alleles $a$ is
\[L(\rho,\xi,\phi,\eta \cd \mathbf{z}, \mathbf{n})=\prod_m\prod_a L_{ma}(z_{ma})\]
where
$$L_{ma} (z_{ma})= \left\{ \begin{array}{cr} g\{z_{ma};\rho D_a(\phi,\xi,\mathbf{n}),\eta\}& \mbox{ if $z_{ma}\geq C$}\\G\{C;\rho D_a(\phi,\xi,\mathbf{n}),\eta\}&\mbox{otherwise,}\end{array}\right.$$
with $g$ and $G$ denoting the gamma density and cumulative distribution function respectively and $D_a$ are the effective allele counts after stutter in \eqref{Ds}. For simplicity of notation, we
have suppressed potential marker and allele dependence of $\rho$,  $\eta$, and $\xi$ in the above formulae.

For a given hypothesis $\hyp$,
 the full likelihood is obtained by summing over all possible
combinations of genotypes $\mathbf{n}$  with probabilities $P(\mathbf{n}\cd \hyp)$ associated with the hypothesis to give
\begin{equation}\label{eq:likely}L(\hyp) = \Pr(E\cd \hyp) = \sum_{\mathbf{n}} L(\rho,\xi,\phi,\eta\cd \mathbf{z},\mathbf{n}) P(\mathbf{n}\cd \hyp).\end{equation}
The number of terms in this sum is huge for a hypothesis which involves several unknown contributors to the mixture, but can be calculated by Bayesian network techniques \citep{graversen:lauritzen:comp:13}; some details are given in \secref{compute} below.

\paragraph{Estimating unknown parameters}
The likelihood function $L(\hyp)$ involves a number of parameters $(\rho,\xi,\phi,\eta)$ which may be completely or partially unknown.

 One way of dealing with this is to make the likelihood as
 large as possible for each of the competing hypotheses and thus calculate
$$\hat L(\hyp) = \sup_{\rho,\xi,\phi,\eta} \sum_{\mathbf{n}} L(\rho,\xi,\phi,\eta\cd \mathbf{z},\mathbf{n}) P(\mathbf{n}\cd \hyp),$$ corresponding to using maximum likelihood estimates for the unknown parameters, but different estimates under the competing hypothesis.  Since the likelihood function \eqref{likely} can be efficiently computed, it is feasible to maximize it using appropriate numerical optimization methods.

Using maximized likelihood ratios preserves the property that the WoE against a suspect $K_s$ in a mixed trace based on our model can never be stronger than what is obtained by a matching single source DNA profile.
To see this, recall that $\pi_s= \Pr(U=K_s)$ denotes the match probability, \ie the probability that a random individual has the genotype of the suspect $K_s$.
As before, let $\hyp_p$ be the prosecution hypothesis, involving $K_s$ as a contributor and $\hyp_d$ be the defence hypothesis, replacing $K_s$ by a
random individual $U$. Let further $\hat\psi_p=(\hat\rho_p,\hat\xi_p,\hat\phi_p,\hat\eta_p)$ be the maximum likelihood estimates of the unknown
parameters under the prosecution hypothesis and $\hat\psi_d$ the estimates under the defence hypothesis. We then have
\begin{eqnarray*}
\mbox{WoE}(K_s\cd \hat\psi_d,\hat\psi_d)&=&
\log_{10}\frac{\Pr(E\cd \hyp_p, \hat \psi_d) }{\Pr(E\cd \hyp_d, \hat \psi_d)}\leq 
\log_{10}\frac{\Pr(E\cd \hyp_p, \hat \psi_p) }{\Pr(E\cd \hyp_d, \hat \psi_d)}\\&=&\mbox{WoE}(K_s\cd \hat\psi_p,\hat\psi_d)\leq \log_{10}\frac{\Pr(E\cd \hyp_p, \hat \psi_p) }{\Pr(E\cd \hyp_d, \hat \psi_p)}\\&=&\mbox{WoE}(K_s\cd \hat\psi_p,\hat\psi_p)\leq
-\log_{10}{\pi_s}= \mbox{WoE}(K_s)_{\max},
\end{eqnarray*}
where the last inequality is obtained as in \eqref{ inequality}.
Thus, also when parameters are estimated, a mixed trace can never give stronger evidence than a high-quality trace from a single source.

\subsection{Computation}
\label{sec:compute}

As mentioned  in \secref{likfun}, the main computational challenge is associated with the marginalisation over
all unknown DNA profiles needed for evaluating the likelihood
function. This is particularly challenging when the number of contributors increases, as the number of terms in \eqref{likely} becomes intractably huge so brute force summation is impossible. To enable efficient computation of the sum, we use an appropriate Bayesian network to represent the genotypes of the unknown contributors and the observations.
As markers are independent  for fixed model
parameters, the model can be represented by a separate Bayesian network for each
marker. Here we shall only briefly describe the basic structure of these networks and refer the reader to  \cite{graversen:lauritzen:comp:13} for further details on the networks and
associated computational issues.

\paragraph{Representation of genotypes}

As customary we assume a reference population in Hardy-Weinberg equilibrium so we can consider the two alleles of an individual chosen at random with allele frequencies $(q_1, \ldots, q_A)$. We recall from \secref{artef} that the distribution of peak height at allele $a$
depends on both $n_{ia}$ and $n_{i,a+1}$ for both known and unknown
contributors $i$. To enable simple computation of the terms in the likelihood function \eqref{likely}, we therefore represent the genotype for contributor
$i$  by a vector of allele counts $(n_{i1}, \ldots,
n_{iA})$. These vectors then follow independent multinomial distributions with $\sum_a n_{ia}=2$.

\begin{figure}[htb]
  \centering
  \newcommand{\myunit}{1em}
  \begin{tikzpicture}[every path/.style={->, >=latex'}]
    
  \tikzstyle{BNnode}=[shape = ellipse, fill = white, draw, font = \scriptsize]
  \tikzstyle{GTmat}=[matrix of math nodes,  execute at empty cell={\node [draw = none, fill = none]{};},
  ampersand replacement=\&, nodes = {BNnode}, row sep = \myunit, column sep =\myunit]
  \tikzstyle{Oarrow}=[]

  \matrix (gt) [GTmat]{
    S_{i1} \& S_{i2} \& S_{i3} \& S_{i4} \& S_{i5} \& S_{i6} \\
    n_{i1} \& n_{i2} \& n_{i3} \& n_{i4} \& n_{i5} \& n_{i6} \\
    [+\myunit]
    O_1 \& O_2 \& O_3 \& O_4 \& O_5 \& O_6 \\
    [+\myunit]
    n_{j1} \& n_{j2} \& n_{j3} \& n_{j4} \& n_{j5} \& n_{j6} \\
    S_{j1} \& S_{j2} \& S_{j3} \& S_{j4} \& S_{j5} \& S_{j6} \\
  };
  
  \foreach \this [evaluate = \this as \prev using int(\this-1)] in {1, ...,6}
  {
    \draw (gt-2-\this) -- (gt-1-\this); 
    \draw (gt-4-\this) -- (gt-5-\this); 
    \draw [Oarrow](gt-2-\this) -- (gt-3-\this); 
    \draw [Oarrow](gt-4-\this) -- (gt-3-\this); 
    
    \ifthenelse{\NOT 1 = \this}{ 
      \draw (gt-1-\prev) -- (gt-1-\this); 
      \draw (gt-1-\prev) -- (gt-2-\this); 
      \draw (gt-5-\prev) -- (gt-5-\this); 
      \draw (gt-5-\prev) -- (gt-4-\this); 
      \draw [Oarrow](gt-2-\this) -- (gt-3-\prev); 
      \draw [Oarrow](gt-4-\this) -- (gt-3-\prev); 
    }
    {};
  }
  \begin{pgfonlayer}{background}
    \node[rectangle, rounded corners = \myunit, draw = none, fill = black!10, fit = (gt-2-1)(gt-2-6)](bg1){};
    \node[rectangle, rounded corners = \myunit, draw = none, fill = black!10, fit = (gt-4-1)(gt-4-6)](bg2){};
    \node[rectangle, rounded corners = \myunit, draw = none, fit = (gt-1-1)(gt-1-6)](bg1a){};
    \node[rectangle, rounded corners = \myunit, draw = none, fit = (gt-5-1)(gt-5-6)](bg2a){};
  \end{pgfonlayer}
  \tikzstyle{tbox}=[font = \scriptsize, rectangle, inner sep = 0, text width = 3cm]
  \node [right = 0.5\myunit of bg1, tbox](p1){Genotype for unknown contributor $i$};
  \node [right = 0.5\myunit of bg2, tbox](p2){Genotype for unknown contributor $j$};
  \node [right = 0.5\myunit of bg1a, tbox](p1a){Partial sums of allele counts};
  \node [right = 0.5\myunit of bg2a, tbox](p2a){Partial sums of allele counts};
\end{tikzpicture}  

\caption{Bayesian network modelling the genotypes of 2 unknown
  contributors $i$ and $j$ and peak height observations for a marker with 6 possible allelic types.}
  \label{fig:gtnet}
\end{figure}
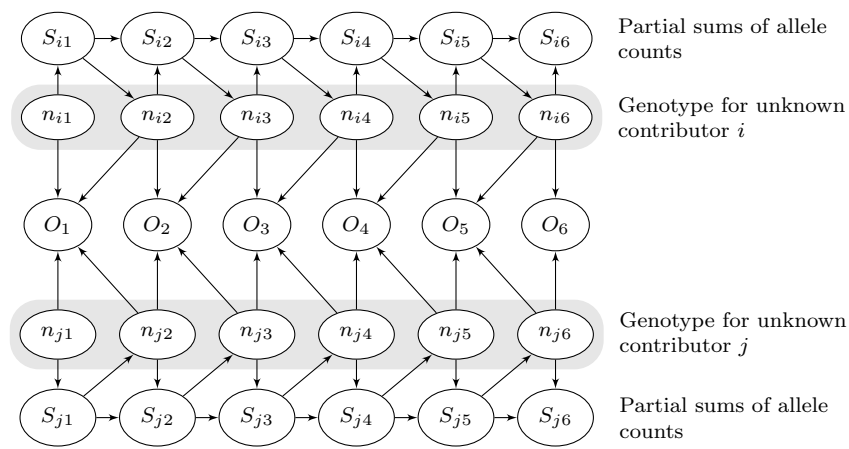

Using properties of the multinomial distribution we can describe this distribution sequentially as follows. 
The number
$n_{i1}$ of the first allelic type follows a binomial distribution $\mathrm{Bin}(2 ,
q_1)$. Denote by $S_{ia} = \sum_{b\le a} n_{ib}$ the number of alleles of
type  1 to $a$ that contributor $i$ possesses.  For any subsequent
allelic type $a+1$, it holds that, conditionally on $S_{ia}$, the number of
alleles of type $a+1$ is independent of the specific allocations
$(n_{i1},\ldots, n_{ia})$ already made.  Furthermore, $n_{i,a+1}$
is also binomially distributed with $n_{i,a+1}|S_{ia} \sim \mathrm{Bin}(2 - S_{ia},
q_{a+1}/\sum_{b\ge a+1} q_b)$.
Thus, adding the partial sums of allele counts to the network  allows the
genotype to be modelled by a Markov structure as displayed in
\figref{gtnet}.

\paragraph{Including peak height information}
The information about peak height observations is incorporated via binary nodes 
$O_a$ in \figref{gtnet}; these represent
whether a peak for allele $a$ has been seen. By suitable specification of the conditional distribution of $O_a$ given its parent nodes, 
conditioning on $O_a$ being TRUE or FALSE will correspond to
conditioning on the observed peak heights $Z_a$;  this enables in particular fast evaluation of the likelihood function \eqref{likely} by a single propagation in the Bayesian network  \citep{graversen:lauritzen:comp:13}.

The computational effort is exponential in the number of unknown contributors. However, the Markov representation of the genotypes themselves ensures the complexity to grow linearly with the number of allelic types $A$, rather than polynomially as in previously used representations \citep{article:Gammamodel,rgc/sll/jm:fsi,Cowell2011202}.

\section{Related literature}
\label{sec:litreview}

Methods for the interpretation of DNA profiles arising from mixtures can be classified into:

\begin{enumerate}[a)]
  \item those mostly based on qualitative information, \ie information
about  allele presence or absence in the DNA mixture;
  \item those that also use quantitative information by taking into account both allele presence and peak intensities.
 \end{enumerate}

We will first discuss  some recent relevant papers that base their model prevalently on allele presence and secondly we describe those   that
also exploit the information in the peak intensities.   A list of publications relating to  DNA mixtures is maintained at  the website of the National Institute of Standards and Technology (NIST)\footnote{\texttt{http://www.cstl.nist.gov/strbase/mixture.htm}}.

\subsection{Methods based on qualitative information} 
Recently, the International Society for Forensic Genetics published  recommendations based on discrete models for forensic analysis of single
source DNA profiles \citep{Gill2012}. They consider a single locus (DNA marker) and at that locus
allow for artefacts such as dropin and dropout, including the probability of their occurrence in the likelihood ratio
computations. Under the independence assumptions of this model, they state that it readily extends to multiple loci and DNA mixtures.
\citet{Gill2012} is largely based on
 \citet{Curran2005} and \citet{balding:buckleton:09}.

  \citet{Curran2005} give a set based method for likelihood ratio calculations which includes multiple contributors
 and the analysis of replicate runs. This was an  improvement on previous guidelines for the analysis of LTDNA profiles where alleles were not reported unless they were
 duplicated in replicate runs \citep{gill2000}. They consider the probability of contamination and dropout, but not of stutter.
 The model is implemented in \citet{gill2007}. For the interpretation  of  LTDNA, \citet{Gill200891} include the probabilities of dropout,
 contamination and stutter; peaks in stutter positions are considered ambiguous alleles and
 included in the calculation. Furthermore, they account for dropin by
   increasing the number of potential contributors to the mixture.
 They also study the robustness of the WoE to misleading evidence in favour of the prosecution.

  \citet{balding:buckleton:09} consider a discrete model for interpreting LTDNA
mixtures where  peaks are classified  as  present, absent, or masking. The set of masking alleles is defined as every peak
above a designated threshold that either corresponds to an allele of a known contributor,  is in a stutter position to that allele or to another peak of sufficient height.  Based on an extension of
this paper \citep{balding:13}, Balding wrote a suite of R functions \texttt{likeLTD}\footnote{\texttt{https://baldingstatisticalgenetics/software/likeltd-r-forensic-dna-r-code}} built for a  range of crime scene DNA profiles, involving complex mixtures, uncertain allele designations, dropin and dropout,
degradation, stutter,  relatedness of alternative possible contributors, as well as replicated runs. Alleles are defined as present, absent, or
uncertain;  dropout is modelled using \citet{Tvedebrink2009,Tvedebrink2012}.  \citet{Haned2011525} estimate dropout probabilities using different logistic models for haploid and
diploid cells in combination with a simulation model of the PCR process \citep{gill2005}.
 Similarly, \citet{HanedSlooten2012} and  \citet{HanedGill2011} adopt a model based on
allele presence that analyses the sensitivity to dropin and dropout parameters. The papers mentioned in this paragraph do not make direct use of peak heights, but peak heights enter implicitly in the process of allele classification.

\subsection{Methods based on quantitative information}  Early attempts to exploit peak intensity information include  \cite{article:Evettdata,article:Perlindata,wang2006least,article:PENDULUM}; and \cite{CurranMCMC}. \citet{article:Gammamodel} introduced  the use of  gamma distributions to  model  peak area variability 
observed in the PCR amplification of DNA in mixtures, and presented a normal-approximation version of the model  in \citet{rgc/sll/jm:fsi}. Both  of these papers carry out calculations using exact propagation algorithms for Bayesian networks, but require
discretization 
of a parameter  representing the relative amounts of DNA from each contributor to the mixture. Both evidential and deconvolution analyses were presented.  The suitability of the gamma distribution was investigated  in
 \citet{Cowell2009}, using the simulation model of  \citet{gill2005}. 

The gamma model of \citet{article:Gammamodel} was extended in \cite{Cowell2011202} to handle  stutter, dropout and silent alleles. Stutter  was represented in Bayesian networks by means of a discretization to allow probability propagation. The authors showed how the Bayesian network representation readily  
enabled the simultaneous analysis of two DNA mixtures that are thought to have contributors in common.
The authors noted that the computational complexity of the model increased significantly with the total number of contributors, limiting the practical  use of their methods to at most four contributors.

\citet{perlin:2011} present a model for DNA mixture interpretation based on a normal model extending the method in \citet{article:Perlindata} in a direction similar to \citet{rgc/sll/jm:fsi}. They use
a fully Bayesian approach where all parameters are given a prior distribution. The model is used both for identification of contributors to a DNA
mixture and for deconvolution. The model does not incorporate stutter, and dropout is accounted for by a background variance parameter. The performance of the model is reported for 16 two-person mixture
samples, however  the performance for mixtures of DNA from more than two people is not reported.

\citet{puch2013} is based on using the gamma distribution for peak heights as in
\citet{article:Gammamodel,Cowell2011202}. Since DNA degrades more the greater the DNA fragment length \citep{Tvedebrink201297}, they use a
correction for this phenomenon. Their methodology is different to the one presented here in that: i) they exploit the common scale parameter for  peak heights to derive Dirichlet distributions for 
relative peak heights; ii) they use a pre-processing step for classifying peaks as possible stutter peaks; iii) they use a more complex dropout model; iv) they discretize the parameter representing the proportion of DNA from each
contributor and use a finer grid on extreme intervals to better capture unbalanced mixtures; v) all alleles not
corresponding to peaks above threshold are lumped together into a single compound allele. They discuss identification of contributors
to the mixture and not  deconvolution. The paper presents examples of two person mixtures and does not analyze replicate runs or multiple
traces.

 \citet{Tvedebrink2010} evaluate the weight of evidence for two person mixtures, using a multivariate normal distribution of peak heights.
 Based on DNA mixtures from controlled experiments they find a linear relationship between peak height and area and between means and variances of
 peak height measurements.
  Controlled experiments are also used for estimating the probability of allelic dropout per locus using logistic regression
   \citep{Tvedebrink2009, Tvedebrink2012}.

Recently, \citet{Taylor2013516} used a log-normal model for the ratio between observed and expected peak heights.

\section{Case analysis}
\label{sec:case}
We proceed to illustrate the methodology developed by applying it to the case
described in \secref{pubcase}. We have used a threshold
of $C=50$ for both traces MC15 and MC18; note that for MC18 the peaks for
alleles $21$ and $25$ at marker FGA are of height $49$ and $39$ and thus below the threshold.
Using a threshold of $C=39$ for this profile makes a negligible
difference to the results.

The computations were all performed with the R-package
\texttt{DNAmixtures} \citep{graversen:package:13} which is available
from \texttt{dnamixtures.r-forge.r-project.org}. The R-package
interfaces the HUGIN API \citep{hugin:api:manual:2012} through the
R-package \texttt{RHugin} \citep{manual:RHugin}. The likelihood
functions are maximized numerically using \texttt{Rsolnp}
\citep{rsolnp,ye:87}. Approximate standard errors of estimates are
based on the inverse Hessian of the likelihood function found by
numerical derivation using \texttt{numDeriv} \citep{numderiv}. The
population allele frequencies used were taken from the US Caucasian
database in \cite{butler:etal:03}. We note that the typical
computation time on a standard laptop for producing the analysis of a
single trace as in \tabref{mlesep} with the current
version of the package is a few minutes.  The computational effort
grows exponentially with the number of unknown contributors, but grows
linearly with the number of alleles and markers
\citep{graversen:lauritzen:comp:13}.

\subsection{Weight of evidence}
\paragraph{Analysis of MC15}
The maximum likelihood estimates and their approximate standard errors under the prosecution
hypothesis $\hyp_p: K_1\amp K_2\amp K_3\amp U$ and defence hypothesis $\hyp_d: K_1\amp K_2\amp U_1\amp U_2$ are given in \tabref{mlesep}.
\begin{table}
\caption{\label{tab:mlesep}Maximum likelihood estimates based on MC15 and MC18 analysed separately.}
{\centering\small
\begin{tabular}{lrrrr|lrrrr}
\multicolumn{5}{c|}{Prosecution hypothesis}&\multicolumn{5}{c}{Defence hypothesis}\\
&\multicolumn{2}{c}{MC15}&\multicolumn{2}{c|}{MC18}&&\multicolumn{2}{c}{MC15}&\multicolumn{2}{c}{MC18}\\
Parameter&Est.&SE&Est.&SE&Parameter&Est.&SE&Est.&SE\\ \hline
$\mu$	&	913	&	35	&	1056	&	39	&	$\mu$	&	914	&	41	&	1056	&	41	\\
$\sigma$	&	0.171	&	0.018	&	0.166	&	0.017	&	$\sigma$	&	0.198	&	0.023	&	0.172	&	0.019	\\
$\xi$	&	0.074	&	0.014	&	0.085	&	0.016	&	$\xi$	&	0.072	&	0.019	&	0.085	&	0.017	\\
$\phi_{K_1}$	&	0.821	&	0.020	&	0.706	&	0.022	&	$\phi_{K_1}$	&	0.798	&	0.028	&	0.698	&	0.026	\\
$\phi_{K_2}$	&	0.047	&	0.014	&	0.091	&	0.016	&	$\phi_{K_2}$	&	0.039	&	0.020	&	0.096	&	0.018	\\
$\phi_{K_3}$	&	0.124	&	0.015	&	0.194	&	0.018	&	$\phi_{U_1}$	&	0.081	&	0.013	&	0.193	&	0.020	\\
$\phi_U$	&	0.008	&	0.019	&	0.009	&	0.018	&	$\phi_{U_2}$	&	0.081	&	0.013	&	0.013	&	0.021	\\
\hline
$\log_{10}\hat L(\hyp_p)$&
\multicolumn{2}{c}{-118.0}&
\multicolumn{2}{c|}{-130.1}&
$\log_{10}\hat L( \hyp_d)$&
\multicolumn{2}{c}{-129.3} &
\multicolumn{2}{c}{-143.4}\\
\hline
\end{tabular}
}
\end{table}
The last unknown contributor is mainly included to allow for potential
spontaneous dropin.  We note that the estimates for the parameters
$(\mu,\sigma,\xi)$ are very similar under the two hypotheses, the
defence hypothesis suggesting a slightly larger coefficient of
variation, both a little less than 20\%, i.e.\ indicating large variability of the peak heights.  The mean stutter proportion
$\xi$ is estimated to just under 7.5\%. The estimates of the
contributor fractions $\phi$ under the prosecution hypothesis agree
well with the estimates $(0.80,0.06,0.12,0.04)$ found in Table 3 of
\cite{Gill200891}.  The resulting WoE is $\log_{10}( LR)=
-118.0-(-129.3) = 11.3$ bans.  For the prosecution hypothesis we
can ignore spontaneous dropin, the standard deviation of $\phi_U$
indicating this may well be zero. Under the defence hypothesis there
are difficulties distinguishing the two unknown contributors in that
their contributor fractions are estimated to the same value. Since the
estimates for $\phi_{U_i}$ are on the boundary of the parameter space,
their standard errors are given for the model which further restricts
them to be equal as $\phi_{U_1}= \phi_{U_2}$.  Removing one unknown
contributor for both hypotheses increases the WoE to $12.1$
bans. This should be compared to the analysis in \cite{Cowell2011202}
which for hypotheses with three contributors gave a WoE around 10
bans. It should also be compared to the upper bound of 14.5 bans
calculated from the match probability for $K_3$.

Thus our model gives considerably stronger WoE than found previously, although much less than would have been possible for a perfect single
source trace. The loss of efficiency compared to a single
source trace is similar to the loss from leaving out, for example,
markers D18 and D19; see \secref{objectives}.

\paragraph{Analysis of MC18}
The maximum likelihood estimates and their approximate standard errors
under the defence and prosecution hypotheses are given in
\tabref{mlesep}.

We note again that the estimates for $(\mu,\sigma,\xi)$ are very
similar under the two hypotheses, the defence hypothesis suggesting a
slightly larger coefficient of variation, but the same order of
magnitude as for MC15. Also here the estimates of the contributor
fractions $\phi$ under the prosecution hypothesis agree well with the
estimates $(0.67,0.11,0.19,0.04)$ in Table 2 of \cite{Gill200891}. In
contrast to MC15, the defence hypothesis identifies an unbalanced
contribution to MC18 from the unknown contributors, the major unknown
contributor representing a fraction similar to that of the defendant
$K_3$ under the prosecution hypothesis. The estimates of the mean
stutter proportions are slightly larger than found for MC15, although
the standard errors suggest that the differences are well within what
could be expected from random variation.  The resulting WoE for this
trace becomes $13.3$ bans, which compares to losing the information
in  marker D18, say. 

Standard deviations for contributor fractions indicate that one unknown contributor can be removed for both hypotheses; this has no essential effect on the 
 WoE which remains at $13.3$ bans. We compare again to
\cite{Cowell2011202} which gave a WoE around 10 bans, again a 
weaker evidential value than what we obtain from our
model.

\paragraph{Combined analysis of MC15 and MC18}

The combined analysis of more than one DNA profile can be very
informative when computing the WoE but especially for deconvolution
of mixed traces, see for example \Secref{deconv}. This is
particularly true when there might be high dropout probabilities in
one or more of the profiles. Multiple profiles with the same
contributors can, for example, be obtained from different parts of a
crime stain, but also from different stains found at the same
scene. In addition, as we shall see, the combined analysis may
considerably improve the precision of parameter estimates.
 
For a combined analysis of the two traces there is a range of
possibilities. The simplest possible combined analysis assumes both
for the prosecution and defence that the unknown contributors to the
two traces are different and unrelated. Under this assumption, the
traces are independent and the joint likelihood ratio is obtained by
multiplying the two likelihood ratios found above to yield
overwhelmingly strong evidence at $24.6$ bans. However, this defence hypothesis is extremely unfavourable for the defence; as
we shall see below, the observed evidence is much more probable under the
assumption that the unknown persons are identical for the two traces, yielding more sensible numbers for the WoE. This highlights the relativity of the WoE to the hypotheses chosen: a very high WoE could in principle be due to the choice of a completely inadequate defence hypothesis.

As the traces are dependent when unknown contributors are shared, a
fresh calculation is needed to find the likelihood ratio. In this
calculation we assume that the two traces have identical mean stutter
proportions $\xi$ and scale parameters $\eta$. This seems reasonable
since these parameters refer to variability in the determination of peak heights rather than to the specific
traces.  In addition we assume that the unknown
contributors are the same for the two profiles, yielding the results
displayed in \tabref{mle1518}.

\begin{table}
\caption{\label{tab:mle1518}Maximum likelihood estimates when information in the traces MC15 and MC18 is combined.}
\centering \small

\begin{tabular}{lrrrr|lrrrr}
&\multicolumn{4}{c}{Prosecution hypothesis}&&\multicolumn{4}{c}{Defence hypothesis}\\
&\multicolumn{2}{c}{MC15}&\multicolumn{2}{c}{MC18}&&\multicolumn{2}{c}{MC15}&\multicolumn{2}{c}{MC18}\\
Par.&Est. &SE&Est.&SE&Par.&Est.&SE&Est.&SE\\
\hline
$\mu$   &       914&    36&  1055  & 38     &$\mu$ &   914&  36& 1055  &39   \\
$\sigma $ &     0.175 & 0.013 &  0.163& 0.012 &    $\sigma$   &   0.178 &0.013&  0.165&0.013\\
$\xi$      &  0.079& 0.011& 0.079&0.011&    $\xi$      &  0.079& 0.011 & 0.079&0.011\\
 $\phi_{K_1}$&  0.822&0.020& 0.705&0.021  &       $\phi_{K_1}$ & 0.820 &0.021&  0.702&0.023\\
  $\phi_{K_2}$& 0.048&0.014 &0.090&0.016    &       $\phi_{K_2}$& 0.049& 0.014& 0.091&0.017\\
  $\phi_{K_3}$ &  0.125& 0.016&0.193  &0.018  &       $\phi_{U_1}$ &  0.123&0.016 & 0.193&0.018\\
$\phi_U$ &  0.006 &0.018& 0.012   &0.017  &     $\phi_{U_2}$& 0.008 &0.019& 0.014&0.018\\
\hline
\multicolumn{2}{l}{$\log_{10}\hat L(\hyp_p)$}&\multicolumn{3}{r|}{-248.2}&\multicolumn{2}{l}{$\log_{10}\hat L(\hyp_d)$}& \multicolumn{3}{r}{-262.3}\\
\hline
\end{tabular}

\end{table}

We note that the standard errors in \tabref{mle1518} are not
dissimilar to previous values, apart from the standard errors of $\hat
\sigma$ and $\hat \xi$, which are reduced considerably by combining
information from the two traces.  The estimates for most parameters
are strikingly similar under the two hypotheses. The fraction
attributed to the defendant $K_3$ under the prosecution hypothesis is
attributed to the major unknown contributor $U_1$ under the defence
hypothesis. The interpretations under the prosecution and defence
hypotheses only disagree on the identity of this contributor.  The
resulting WoE for the combination of traces becomes $14.1$ bans.
This evidence has strength similar to that of a single source trace
with a perfect match to the defendant $K_3$ which would yield a WoE of
$14.5$ bans. 
Thus the loss of evidential efficiency $\mbox{WL}(E\cd K_3)$  against $K_3$ is about $0.4$ bans.
As for the single trace analyses one unknown contributor is
redundant and could be removed from the analysis, leaving the WoE unchanged at
$14.1$ bans. \cite{Cowell2011202} gave a WoE of $8.9$ bans
using their model.

Above we have analysed the combined traces under the assumption that the unknown contributors to the two traces are the same. To assess whether this assumption is reasonable, we can compare the maximized likelihoods to those allowing  possibly different unknown contributors under the two hypotheses. None of the associated reductions in likelihood are statistically significant. Due to the many one-sided constraints on the parameters, the $p$-values are not easily calculated \citep{elbarmi:dykstra:99}, but can in the present analyses be shown to be at least $0.19$; hence there is no reason to assume that the contributors are different.

We also note from \tabref{mle1518} that the mixture proportions
$\phi$  appear to be similar for the two traces, although he standard deviations indicate that they are not identical;  indeed a likelihood ratio
test for their identity is rejected under both hypotheses with  $p$-values around 0.0003. 

For curiosity we also analysed MC15 and MC18 using
\texttt{likeLTD} obtaining a WoE of $8.8$ bans. This number
is not directly comparable to our analysis above as it refers to a
slightly different database of allele frequencies and a coancestry
parameter $F_{ST}=0.02$ is applied together with a sampling adjustment, resulting in a larger match probability for $K_3$ and thus
a smaller maximal WoE of $13.2$ bans. Furthermore, \texttt{likeLTD}
assumes that the contributor fractions are identical for the two
traces since they are treated as replicate runs of the same 
trace. 

To compare with the analysis obtained by \texttt{likeLTD} we have made an analysis with our model assuming equal contributor fractions, using the same database, sampling adjustment, and coancestry parameter. This correction can readily be made as it only involves appropriate modification of the allele frequencies \citep{balding:13}. Using a likelihood function based only on observed peak presence or
absence, plugging in parameter estimates based on the
observed peak heights, we get a WoE of $10.0$ bans. The corresponding
WoE based on observed peak heights is $12.7$ bans.
It seems fair to conclude that properly exploiting the information in
peak heights gives a more efficient analysis of the traces than what
can be obtained from allele classifications alone; this could be
important for the analysis of mixtures more complex than those
analysed here.

\subsection{Mixture deconvolution}
\label{sec:deconv}
For mixture deconvolution we consider the traces jointly under the defence hypothesis modified by removing one unknown contributor and identify the five most probable full genotypes of the
unknown individual $U$. For most markers, these genotypes share that of the defendant $K_3$; indeed the only variations are on markers D16, D18,
and      D19, where the most probable configurations are displayed in \tabref{mix1518deconv}. The markers not represented in the table share the
genotype of the defendant for all five most probable combinations. 
\begin{table}
\caption{\label{tab:mix1518deconv}Most probable genotypes of the unknown contributor $U$ under the  hypothesis  $K_1\amp  K_2 \amp  U$ when information in the traces MC15 and MC18 is combined.}
\centering \small
\begin{tabular}{lcccr}
&\multicolumn{3}{c}{Marker}&\\
\cline{2-4}
 Rank&{D16}&{D18}&{D19}& Profile probability\\
\hline
1   &     $(11,13)$ &      $(12,16) $   & $(14,15)$&   0.436\\
2   &      $(11,13) $ &      $(12,14) $   & $(14,15)$&   0.129\\
3   &      $(11,13) $ &      $(12,16) $   & $(13,15)$&   0.070\\
4   &      $(11,13) $ &      $(12,16) $   & $(13,14)$&   0.052\\
5   &      $(12,13) $ &      $(12,16) $   & $(14,15)$&   0.047\\
\hline
Total&&&&0.734\\
\hline
\end{tabular}
\end{table}
We  note that the profile with the top rank is indeed that of the defendant, and the evidence gives a probability to the unknown person having precisely this profile of $44\%$. 
The probability that the true profile of the unknown contributor is among the five genotypes listed is $73\%$.

 If we make a deconvolution of the single traces, the identification of the genotype of $U$ is less clear. 
 The distribution of the unknown profile becomes much more diffuse reflected by a drastic reduction in the probability of the most probable profiles: in the case of MC15 to $0.2\%$ and for MC18 to $11\%$. 
  The probability that the true profile of the unknown contributor is among the five most probable profiles is reduced to $1\%$ and $39\%$ compared to $73\%$ for the traces combined.  Also, for the single trace deconvolutions, the profile of the defendant is ranked fourth for both traces rather than first. Thus it is apparent that   combining trace information for deconvolution purposes can be of considerable value.

\subsection{Interpreting artefacts}
\label{sec:artefacts}

 Our model does not impose at the outset that a specific peak or allele is due to stutter, or has dropped out, or is silent. One of the features of \texttt{DNAmixtures} \citep{graversen:package:13} is to produce Bayesian networks for each marker with peak height evidence propagated. Thanks to the flexibility of Bayesian networks we can then elaborate these so as to explicitly represent artefacts. 
 
We can also make simple modifications of the model to allow for the possible presence of silent alleles. 
In this way, we can answer queries like: what is the probability that an observed peak is due only to stutter? that
a specific allele has dropped out? that an unknown contributor possesses a silent allele?

\paragraph{Identifying stutter and dropout}

 To enable a precise analysis we shall say that an observed peak at $a$ is \emph{due to stutter}  if no contributors possess
 allele $a$.  Similarly we say that an allele $a$ has \emph{dropped out} if at
 least one contributor possesses allele $a$, but no peak has been observed above threshold, \ie $Z_a=0$. 
 
To investigate such events we introduce variables $Y_a$ which explicitly represent the presence or absence of the allele $a$ as follows
\[ Y_a=\begin{cases} 1& \mbox{if  $\sum_i n_{ia}>0$} \\ 0 & \mbox{if $\sum_i n_{ia}=0$}.\end{cases}\]
These variables can readily be included in the network as nodes with parents $n_{ia}$ and
probability propagation  would then yield
$P(Y_{a}=1\cd \mathbf{z})$ and $P(Y_{a}=0\cd \mathbf{z})$.
For an allele $a$ where a peak has been observed, the probability that the peak is due to stutter alone is now $P(\mbox{stutter}\cd \mathbf{z})=P(Y_{a}=0\cd \mathbf{z})$, whereas if no peak has been observed at $a$, the probability that the allele $a$ has dropped out is $P(\mbox{dropout}\cd \mathbf{z})=P(Y_{a}=1\cd \mathbf{z})$. A pre-classification of some alleles as, say, stutter or dropout corresponds to conditioning on specific values of $Y_a$; the consequences of this conditioning can be obtained by a single probability propagation. A revised mixture analysis can then be performed conditionally on the pre-classification, should this be desired. We note that this form for preprocessing of data would still be consistent with our general model specification.

We now revisit  the defence hypothesis $\hyp_d: K_1\amp K_2\amp U_1\amp U_2$
using the parameter estimates in \tabref{mle1518} and obtain the results in
\tabref{artefacts}.
\begin{table}
  \caption{\label{tab:artefacts}Posterior probabilities for stutter and
   dropout.}
  \centering
  \begin{tabular}{ccrrrr}
    Trace & Marker &Allele $a$ & $z_a$ & $P(\mbox{stutter}\cd \mathbf{z})$ & $P(\mbox{dropout}\cd \mathbf{z})$\\ 
    \hline
    MC18 & D2 &  15 & 0 &  & 0.003 \\ 
    &&  16 & 189 & 0 &  \\ 
    &&  17 & 171 & 0 &  \\ 
    &&  18 & 0 &  & 0.156 \\ 
    &&  19 & 0 &  & 0.219 \\ 
    &&  20 & 0 &  & 0.274 \\ 
    &&  21 & 0 &  & 0.083 \\ 
    &&  22 & 55 & 0.927 &  \\ 
    &&  23 & 638 & 0 &  \\ 
    &&  24 & 673 & 0 &  \\ 
    &&  25 & 0 &  & 0.178 \\ 
    &&  26 & 0 &  & 0.060 \\ 
    &&  27 & 0 &  & 0.003 \\ 
    \hline
    MC15 & TH01 &  5 & 0 &  & 0.004 \\ 
    &&  6 & 0 &  & 0.328 \\ 
    &&  7 & 727 & 0 &  \\ 
    &&  8 & 625 & 0 &  \\ 
    &&  9 & 0 &  & 1.000 \\ 
    &&  9.3 & 165 & 0 &  \\ 
    &&  10 & 0 &  & 0.018 \\ 
    &&  11 & 0 &  & 0.004 \\ 
    \hline
  \end{tabular}
\end{table}
Conforming with the remarks in \secref{pubcase} we note that the peak at allele 22 for D2 has high probability (93\%) of being a stutter peak. Similarly, we note that the allele 9 for marker TH01 has certainly dropped out, consistent with the fact that $K_2$ possesses this allele. We also note, for example, that it is rather probable (33\%) that allele 6 for marker TH01 is actually present in the mixture despite the fact that neither of $K_1$ and $K_2$ have that allele represented in their genotype and no peak has been observed. This is partly due to this allele having a  high frequency (23\%) in the reference population; but, since no peak has been observed, the fraction of DNA from of an unknown contributor possessing this allele is likely to be very small.

Unsurprisingly, the combined analysis of two traces gives more informative results on the analysis of artefacts than what emerges from separate analyses of the two traces. We refrain from reporting the separate analyses here.

\begin{figure}[tb]
  \begin{center}
    \newcommand{\myunit}{1em}
  \begin{tikzpicture}[every path/.style={->, >=latex'}]
    
  \tikzstyle{BNnode}=[shape = ellipse, fill = white, draw, font = \scriptsize]
  \tikzstyle{GTmat}=[matrix of math nodes,  execute at empty cell={\node [draw = none, fill = none]{};},
  ampersand replacement=\&, nodes = {BNnode}, row sep = \myunit, column sep = \myunit]
  \matrix (gt) [GTmat]{
    S_{i1} \& S_{i2} \& S_{i3} \& S_{i4} \& S_{i5} \& S_{i6} \\
    n_{i1} \& n_{i2} \& n_{i3} \& n_{i4} \& n_{i5} \& n_{i6} \\
    [+\myunit]
    O_1 \& O_2 \& O_3 \& O_4 \& O_5 \& O_6 \\
  };

  \foreach \this [evaluate = \this as \prev using int(\this-1)] in {1, ...,6}
  {
    \draw (gt-2-\this) -- (gt-1-\this); 
    \draw (gt-2-\this) -- (gt-3-\this); 
    
    \ifthenelse{\NOT 1 = \this}{ 
      \draw (gt-1-\prev) -- (gt-1-\this); 
      \draw (gt-1-\prev) -- (gt-2-\this); 
      \draw (gt-2-\this) -- (gt-3-\prev); 
    }
    {};
  }

  \node [BNnode, left = \myunit of gt-1-1](s-i-0){$S_{i0}$};
  \node [BNnode, left = \myunit of gt-2-1](n-i-0){$n_{i0}$};
  \draw (s-i-0)--(gt-1-1);
  \draw (s-i-0)--(gt-2-1);
  \draw (n-i-0)--(s-i-0);

  \begin{pgfonlayer}{background}
    \node[rectangle, rounded corners = \myunit, draw = none, fill = black!10, fit = (gt-1-1)(gt-2-6)](bg1){};
  \end{pgfonlayer}
\end{tikzpicture}  

  \caption{Modified network allowing for silent alleles. The nodes in the shaded area are those in the original genotype representation in \figref{gtnet}.}
  \label{fig:silent}
  \end{center}
\end{figure}
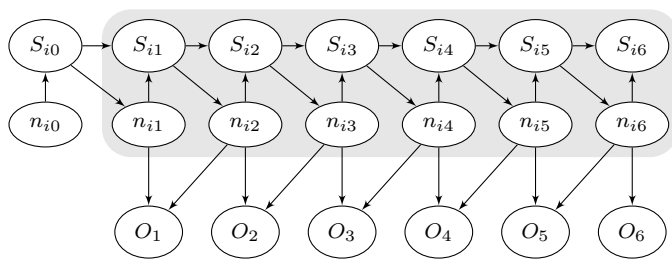

\paragraph{Silent alleles}
 The possibility that the unknown contributors might have a silent
 allele is incorporated in the model by adding an extra allelic type
 $0$ (silent) to the genotype representations as displayed in
 \figref{silent}.
We now let $n_{i0} \sim \mbox{Bin}(2, q_0)$, $S_{i0}=n_{i0}$, $S_{i1}
= n_{i1}+S_{i0}$, and further $n_{i1}\cd S_{i0} \sim
\mbox{Bin}(2-S_{i0}, q_1)$, where $q_0$ is the probability of an
allele being silent and $q_1$ indicates the database allele frequency.
Since no peak can be observed for the silent allele there is no observational
node $O_0$.
This corresponds to a modification of the basic model where the gene
frequencies $q_a, a>0 $ are interpreted as the relative frequencies of
non-silent alleles, so the probability that a random allele is of type
$a$ is equal to $(1-q_0)q_a$ if $a>0$.

\subsection{Single source analysis}
\label{sec:single}

Although our model is developed for the analysis of mixtures, it can also be useful in the analysis of a DNA profile from  a single source; in particular as the peak heights can be informative about the presence of  silent alleles. What appears to be a homozygous genotype at some marker may not be so; an alternative explanation is that we 
see only one allele of a heterozygous genotype, the other allele being silent. We would expect this to be reflected in a peak height that is smaller than expected and it will clearly affect the evidential
interpretation of DNA profiles.

\Tabref{singlesource} shows an excerpt of a DNA profile of a single individual $J$, having apparent homozygous genotypes on markers D8, CSF1PO and FGA.
\begin{table}
\caption{Excerpt of a DNA profile with peak heights of a single individual. \label{tab:singlesource}}
\centering\small 
\begin{tabular}{lcc}
Marker& Allele& Height \\
\hline
   D8 &         13 &       4364 \\
    D21 &         28 &       2646 \\
           &         29 &       2490 \\
    CSF1PO &         11 &       2695 \\
   D3 &         14 &       2249 \\
           &         16 &       2205 \\
      TH01 &          6 &       1268 \\
           &        9.3 &       1294 \\
      TPOX &          8 &       1394 \\
       &         11 &       1526 \\
      AMEL &          X &       4289 \\
    D5 &         11 &       2053 \\
     &         13 &       1827 \\
       FGA &         23 &       2444 \\
\hline
\end{tabular}
\end{table}
Note that the peak heights are rather large, as is customary in single source data with a large amount of DNA.

\begin{figure}[htb]
\begin{center}
\includegraphics{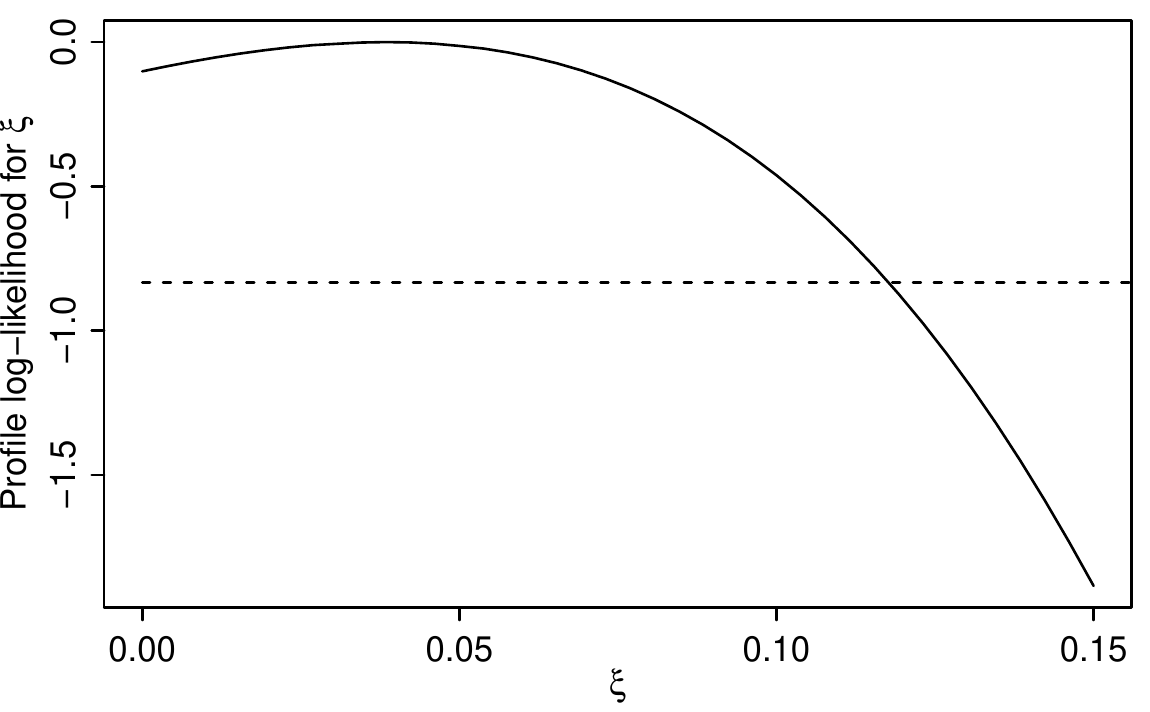}
\end{center}
\caption{\label{fig:profilelik}Profile log-likelihood $\max_{\mu,\sigma}\log_{10} L(\mu,\sigma,\xi)-\log_{10}L(\hat \mu,\hat\sigma,\hat\xi)$ for $\xi$ for the single source trace thresholded at $C=500$. 
Values of $\xi$ with log-likelihood above the horizontal line constitute an approximate 95\% confidence interval.}
\end{figure}

Assuming that the DNA profile in \tabref{singlesource} came from two unknown contributors and 
setting the threshold $C=50$,
we obtain estimates for the fraction of DNA  from  each unknown contributor of  $(\hat\phi_1,\hat\phi_2)=(1,0)$ which clearly
indicate that the DNA trace comes from a single individual.
The estimates of the other parameters are  $\hat{\mu} = 1806$, $\hat{\sigma}= 0.290$, $\hat{\xi}=0$. The combination of a high mean peak height $\hat{\mu} = 1806$ and a vanishing stutter proportion points to the data having been preprocessed so that peaks classified in the laboratory as stutter have been removed.
If we instead set the threshold at $C=500$ to ensure that the removed stutter peaks are well below threshold and assume that the data are from a single individual,   we get the estimates  $\hat{\mu} = 1854$, $\hat{\sigma}= 0.278$ and $\hat{\xi}= 0.038$. Clearly, there is very little information about the stutter proportion when stutter peaks have been removed, resulting in a very flat profile likelihood for the stutter proportion as shown in \figref{profilelik}; indeed a 95\% confidence interval for $\xi$ would be ranging from 0 to 11.8\%.

\begin{figure}[htb]
  \centering
  \includegraphics{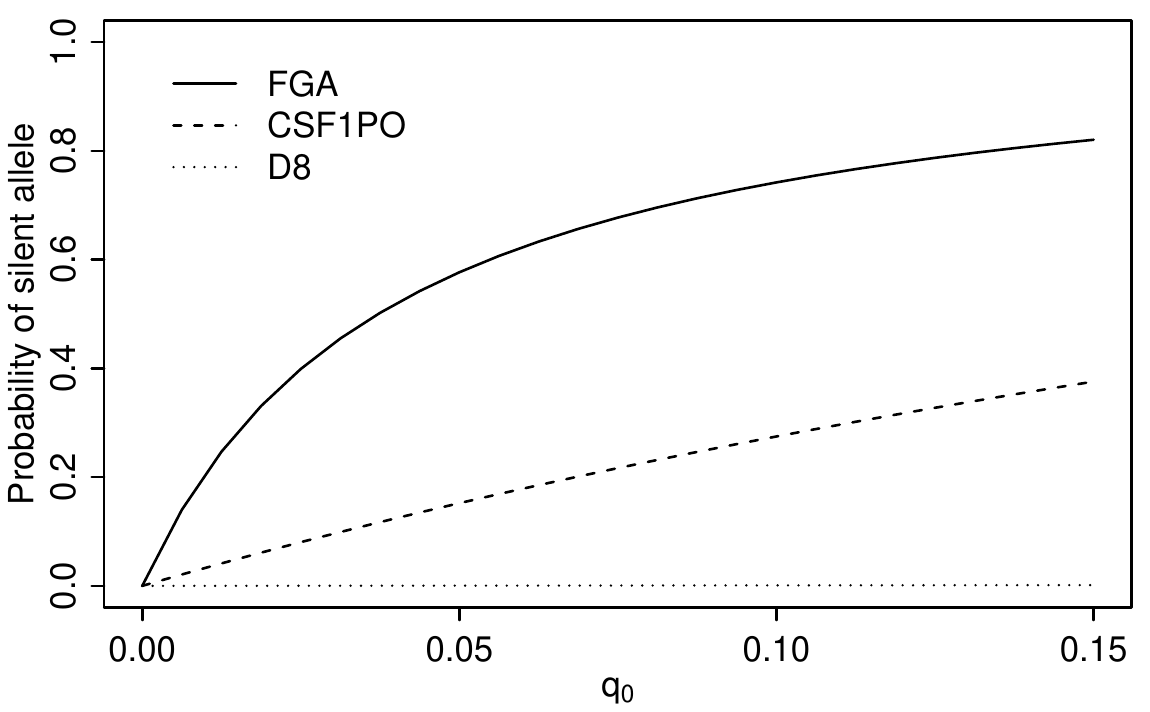}
  \caption{Posterior probabilities that the individual has a silent allele as a function of the apriori probability $q_0$ of having a silent allele.}
  \label{fig:silentprob}
\end{figure}

In \figref{silentprob} we see the posterior probabilities that $J$ has a
silent allele for the markers showing apparent homozygosity: D8,
CSF1PO, and FGA.
As might be expected, for D8 where the peak height is relatively high, the
probability of a silent allele is almost zero. Markers CSF1PO and FGA
have peak heights around 60\% and 56\% that of D8, and correspondingly
higher probabilities of $J$ possessing a silent
allele. 

 However,  care should be taken with interpreting this result as  FGA and CSF1PO often have a lower peak height than other markers, being at the end of the dye panel and thus corresponding to a longer fragment length, see also \secref{extension} below. Evidence calculations and deconvolutions may be less sensitive to this fact, but it could be of crucial importance for the identification of silent alleles.

\section{Discussion}\label{sec:discuss}
On the previous pages we have described a simple and versatile model for description of mixed  DNA profiles and demonstrated how it could be applied to case analysis. We should emphasize that all our computations are direct consequences of our model and exact apart from the use of deterministic numerical optimization methods. Clearly, the model itself represents an approximation to reality, as any model will, but we believe it is a virtue that any further analysis --- whether this be computation of likelihood ratios, deconvolution, or mixture interpretation --- does not need additional approximations, modifications of the model, nor any \emph{ad hoc} heuristics. Also, our method does not rely on a subjective pre-processing of the DNA profile with a manual or automatic allele calling that may introduce an additional source of error. 

In addition to the case discussed in this article, we have tried our methodology on a number of real challenging cases, including difficult LTDNA mixtures, with very sensible and robust results and believe that the methodology is now sufficiently developed for use in proper case work.  However, for any methodology there is room for improvement and below we shall briefly discuss some issues worthy of further attention. 

We note in passing that the model could potentially also be useful in disputed paternity cases or other types of pedigree analysis where the DNA of some of the actors is available only in small quantities as in LTDNA, or is degraded.

\subsection{Number of contributors}
One issue is associated with determining the number of unknown contributors to the trace. The traces considered need at least three unknown contributors to be well explained. We have in the previous analyses introduced an additional unknown contributor to allow for spontaneous dropin, but showed that such an additional contributor had most likely only contributed tiny amounts of DNA to the mixture, if anything at all. 
Typically, a heuristic argument is used where additional contributors are not added unless they appeared to be necessary for explaining the DNA profile in question. 
However, \cite{buckleton2007} and \cite{Biedermann2012689} point out the risks in  using this kind of approach.

\begin{figure}[htb]
\begin{center}
{ \includegraphics{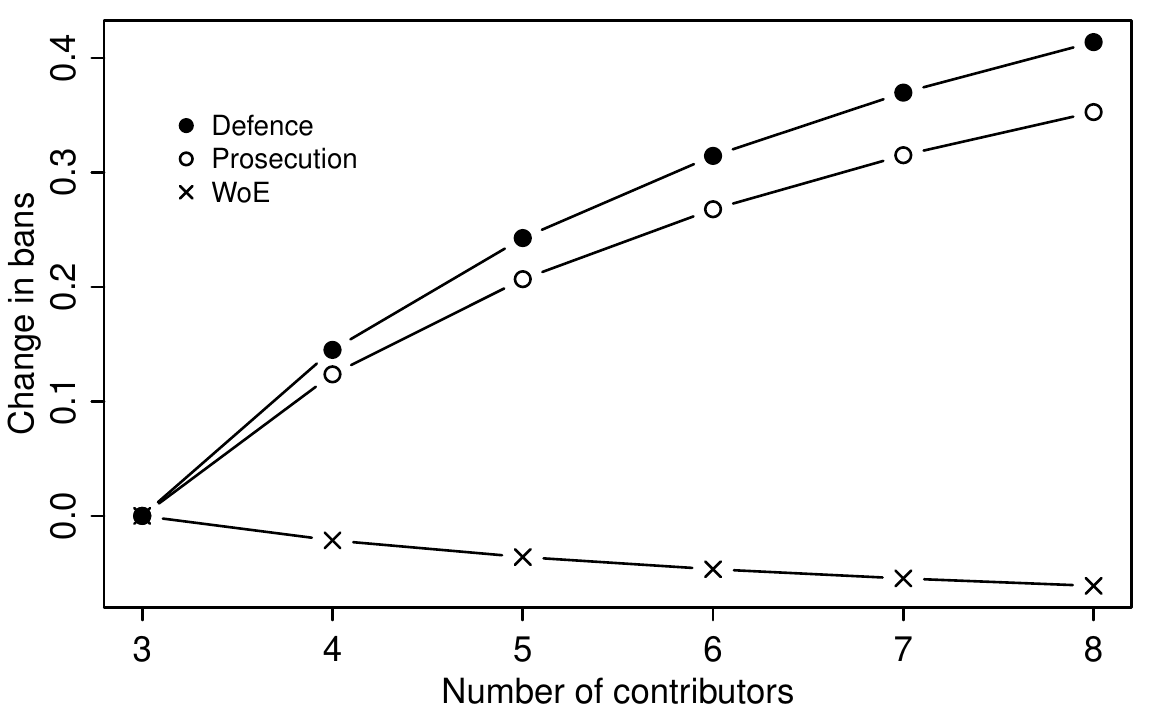}}
\end{center}
\caption{\label{fig:contrib}The maximized log-likelihood functions based on combining traces MC15 and MC18 for varying number of contributors and hypotheses, relative to their value for three contributors and displayed in bans.}
\end{figure}

As we operate with maximized likelihoods under hypotheses that there is at most a fixed number $k$ of contributors, it is obvious that the maximized likelihood will increase --- or at least not decrease --- with the number of unknown contributors,  both for the defence and prosecution. This follows from the fact that the hypothesis of at most $k$ contributors is a sub-hypothesis of that with $k+1$ contributors, so we are maximizing over a larger space when increasing $k$. The phenomenon is illustrated in \figref{contrib}, which shows the maximized likelihoods and corresponding ratios for up to eight contributors, \ie six unknown contributors for the defence hypothesis.
The likelihoods and ratios are normalized relative to what was obtained for three contributors. The likelihood does indeed increase and it increases more for the defence than for the prosecution, so the likelihood ratio decreases with the number of contributors. The change is tiny: for eight contributors the WoE has decreased to $14$ bans from $14.1$ bans  for three contributors,  hardly of any significance. 
Even if the defence is allowed eight contributors and the prosecution stays with three, we get $\mbox{WoE}=13.7$, which is still close to what would be obtained for a perfect match on nine markers and in most cases would imply identification of $K_3$ as a contributor beyond reasonable doubt.

Inspection of the maximum likelihood estimates reveals that the five unknown contributors with smallest amount have between them supplied less than 3.7\% of the total amount of DNA shared in equal proportions, \ie each about 0.75\% of the total amount of DNA.  Bearing in mind that the total solution of DNA contains about 100 human cells, this means that effectively all of their alleles (which are not shared with the three major contributors or in stutter position to those) have not been observed. The unknown contributors just serve the purpose of explaining minor peak imbalances and seem irrelevant for the interpretation of the mixture.  

We believe that the phenomenon of ever increasing likelihoods is an artefact of using direct maximum likelihood without  a penalty for complexity for estimating the unknown parameters. There are various ways of overcoming this problem: for example, one could use a convention that unknown contributors are only added if their presence is statistically significant at some fixed level; one could maximize a likelihood that is penalized for having many contributors; or one could postulate a uniform prior distribution for the unknown fractions $\phi$ and use an integrated rather than maximized likelihood.  All of these would be acceptable but will be essentially \emph{ad hoc} so would need to be established by an agreed convention. We point out that the methodology for bounding the number of contributors in \cite{sll/jm:bound} would not be applicable as this strongly exploits the assumption that no allele has dropped out; the problem in our case is that one could have an infinite number of contributors with so little DNA that  their alleles of their alleles have only been observed through minor perturbations of the peak heights. 

\subsection{Extensions and modifications}\label{sec:extension}
The analyses presented in this paper have been made with the simplest possible variant of the model with sensible and rather robust results. However, there are some obvious potential improvements to be made.
\paragraph{Marker dependence of parameters}
We have assumed that the parameters $(\mu, \sigma, \xi)$ are identical for all markers. This clearly contradicts findings in the literature \citep{book:Butler,puch2013} which indicate that both mean peak heights and stutter proportions depend on marker, fragment length, and may also be different for different dye panels in the EPG. Variation in these parameters can easily be incorporated in the model. However, as this could drastically increase the number of parameters, it would be best to do so by using prior information from laboratory experiments, for example by scaling peak heights as indicated in \cite{puch2013} or by adding prior information by penalizing the likelihood function to ensure that parameter estimates are near those expected.  
\paragraph{Degradation} For strongly degraded DNA samples it is necessary to adjust the model by correcting for fragment length. From the analysis in \cite{Tvedebrink201297} it seems that the simplest and most natural way to do so would be to assume the parameter $\rho$ in the gamma distribution --- being proportional to the amount of DNA --- to depend exponentially on the fragment length $\lambda$ as
$$\log \rho = \delta \lambda + \zeta.$$
This would just add one additional parameter $\delta$ representing the level of degradation of DNA to the model and would not in itself create specific difficulties for the methodology, although the complexity of the maximum likelihood estimation clearly increases whenever new parameters are added.
\paragraph{Dropout}
We have chosen a simple approach to modelling dropout by taking the consequence of our gamma model and using the probabilities for being below thresholds as dropout probabilities. The dropout model of \cite{Tvedebrink2009,Tvedebrink2012} can be seen as using a logistic distribution rather than a gamma distribution for this purpose. The choice of the logistic model seems primarily motivated by the ability to analyse dropout data using standard software and may not in itself have a firm theoretical foundation. In any case it would be interesting to see whether any of the gamma and logistic models would best describe the experimental data behind \cite{Tvedebrink2009}.  We note that since \cite{Tvedebrink2009} uses average peak height as a proxy for the amount of DNA and the relation between peak height and amount may be different in LTDNA analysis, their parameter estimates may not be directly suitable for this purpose but need modification, as described  in \cite{Tvedebrink2012}. Our estimation process would naturally fit the parameters to any specific trace in question.
\subsection{Other issues}
\begin{figure}[htb]
\begin{center}
{ \includegraphics{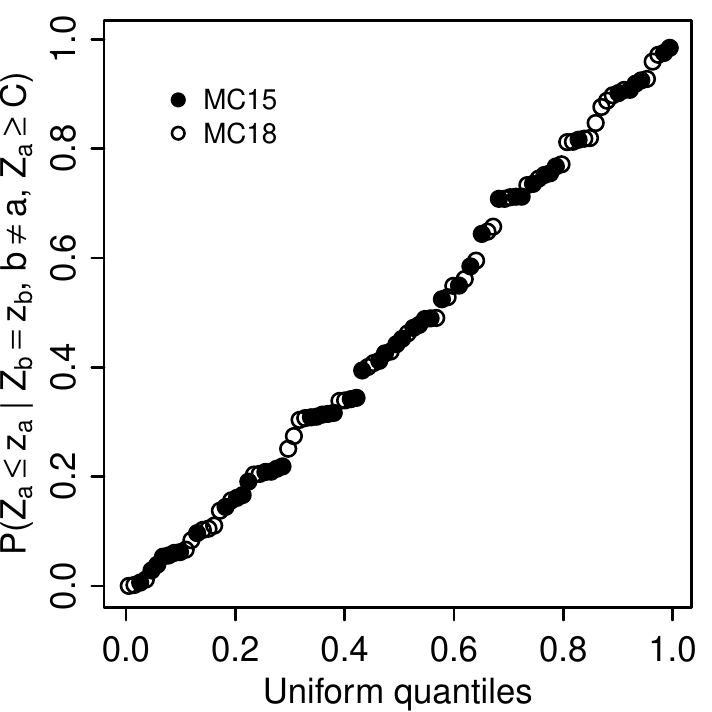}}
\end{center}
\caption{\label{fig:qqplot}Conditional probability transform of observed peak heights for MC15 and MC18 vs.\ uniform quantiles.}
\end{figure}

\paragraph{Model criticism}
An important point that we have not touched upon in the previous analyses is the need to validate the model used for any given dataset. We find that it is not enough to compare the likelihoods for two competing hypothesis if neither of them can be demonstrated to give a plausible explanation of the data at hand. Promising graphical methods for systematic model validation and criticism  based on its internal predictions and residual analysis are important and under development \citep{graversen:lauritzen:comp:13}. These methods exploit that the computational structure of a Bayesian network naturally enables fast prediction of peak heights based on partial information from the profiles. An example is given in \figref{qqplot} which shows the conditional probability transform of peak heights for MC15 and MC18 under the prosecution hypothesis against quantiles of the uniform distribution, indicating an excellent fit of the gamma model. 

\paragraph{Uncertain allele frequencies}
We have in the previous analysis treated the allele frequencies as fixed and known, although in principle they also are parameters which should be estimated. Using allele counts from databases  directly as frequencies corresponds to estimating allele frequencies by maximum likelihood and using the estimated frequencies as if they were known. This conforms with the way other parameters are treated in the present paper, but ignores the inherent uncertainty associated with the frequencies, see also \cite{curran:etal:02} and \cite{curran:05} for a discussion of this aspect. It would be valuable to develop a scheme similar to that in \cite{green:mortera:09} for incorporating both this uncertainty and kinship corrections based on the possibility of alleles that are \emph{identical by descent} (IBD) and we hope to address this in future work. As noted in \cite{green:mortera:09}, the simple kinship correction used in \cite{balding:13} and in our case analysis ignores the dependence between markers created by IBD or by population heterogeneity, which may generally still lead to an overstatement of the WoE.  
\paragraph{Estimation uncertainty} We note that in addition to the sampling uncertainty associated with the allele frequencies, we also have statistical uncertainty associated with estimation of other model parameters. Although we have indicated the  uncertainty of the latter by giving approximate standard errors based on the Hessian of the log-likelihood function, we have neither made a detailed analysis that justifies this, nor have we incorporated this additional uncertainty into our analysis. However it would be feasible to do so along the lines described in \cite{graversen:lauritzen:jas:13} and this will be further developed in the future.

\section*{Acknowledgements}
This research was partially supported by Universit\`{a} Roma Tre ``PESFoG Fondi Internazionalizzazione'' Grant. We are indebted to Marina Dobosz for
providing the EPGs in \figref{epg} and Francesca Scarnicci  
for providing the data for \secref{single}. In addition we have benefited from Ate Kloosterman's explanation on details of the PCR process, and discussions with other participants of the PESFoG workshop in ``Villino Volterra'' Ariccia 2012.

\end{document}